%% file: main.tex
\documentclass[notitlepage, 12pt]{article}

\usepackage{amssymb,apacite, amsmath,amsfonts,eurosym,geometry,ulem,graphicx,caption,color,setspace,sectsty,comment,footmisc,caption,natbib,pdflscape,array,hyperref,subcaption,threeparttable,threeparttablex, amsthm, rotating,chngcntr,pdflscape,subfloat, pifont, mathtools, csquotes, longtable, listings, xcolor}

\definecolor{codegreen}{rgb}{0,0.6,0}
\definecolor{codegray}{rgb}{0.5,0.5,0.5}
\definecolor{codepurple}{rgb}{0.58,0,0.82}
\definecolor{backcolour}{rgb}{0.95,0.95,0.92}

\lstdefinestyle{mystyle}{
    backgroundcolor=\color{backcolour},   
    commentstyle=\color{codegreen},
    keywordstyle=\color{magenta},
    numberstyle=\tiny\color{codegray},
    stringstyle=\color{codepurple},
    basicstyle=\ttfamily\footnotesize,
    breakatwhitespace=false,         
    breaklines=true,                 
    captionpos=b,                    
    keepspaces=true,                 
    numbers=left,                    
    numbersep=5pt,                  
    showspaces=false,                
    showstringspaces=false,
    showtabs=false,                  
    tabsize=2
}

\newcolumntype{L}{>{\centering\arraybackslash}m{1cm}}
\newcolumntype{C}[1]{>{\centering\let\newline\\arraybackslash\hspace{0pt}}m{#1}}
\newcolumntype{R}[1]{>{\raggedleft\let\newline\\arraybackslash\hspace{0pt}}m{#1}}

\begin{document}

\begin{titlepage}
\title{EmTract: Extracting Emotions from Social Media\thanks{I am extremely grateful to Stefania Albanesi, Lee Dokyun, Sera Linardi, and Zhi Da for their continued guidance and support throughout this project. Special thanks to StockTwits for sharing their data. This research was supported in part by the University of Pittsburgh Center for Research Computing through the resources provided.}}

\author{Domonkos F. Vamossy\thanks{Department of Economics, University of Pittsburgh, d.vamossy@pitt.edu.}
\and
Rolf P. Skog\thanks{Independent Researcher, Seattle, WA, USA, r.skog1996@gmail.com.}
}
\date{\today}
\maketitle

\noindent

\begin{abstract}
	\singlespacing
We develop an open-source tool (EmTract) that extracts emotions from social media text tailed for financial context. To do so, we annotate ten thousand short messages from a financial social media platform (StockTwits) and combine it with open-source emotion data. We then use a pre-tuned NLP model, DistilBERT, augment its embedding space by including 4,861 tokens (emojis and emoticons), and then fit it first on the open-source emotion data, then transfer it to our annotated financial social media data. Our model outperforms competing open-source state-of-the-art emotion classifiers, such as Emotion English DistilRoBERTa-base on both human and chatGPT annotated data. Compared to dictionary based methods, our methodology has three main advantages for research in finance. First, our model is tailored to financial social media text; second, it incorporates key aspects of social media data, such as non-standard phrases, emojis, and emoticons; and third, it operates by sequentially learning a latent representation that includes features such as word order, word usage, and local context. Using EmTract, we explore the relationship between investor emotions expressed on social media and asset prices. We show that firm-specific investor emotions are predictive of daily price movements. Our findings show that emotions and market dynamics are closely related, and we provide a tool to help study the role emotions play in financial markets.

\noindent \\
\noindent\textbf{Keywords:} Deep Learning; Investor Emotions; Text Analysis; Social Media; Return Predictability. \\
\vspace{0in}\\
\noindent\textbf{JEL Codes: G41; L82.} \\
 \bigskip
\end{abstract}

\setcounter{page}{0}
\thispagestyle{empty}
\end{titlepage}
\pagebreak \newpage

\doublespacing

\clearpage 

\input{Sections/Introduction}

\input{Sections/Related}

\input{Sections/Validation}

\input{Sections/Primary}

\input{Sections/Conclusion}

\input{Sections/References}

\input{Sections/Appendix}

\input{Sections/OnlineAppendix.tex}

\end{document}

%% file: Sections/Introduction.tex
\section{Introduction} \label{sec:introduction}
The connection between financial markets and emotions is well-known. Most have heard the famous adage from Warren Buffett: ``Be fearful when others are greedy, and greedy when others are fearful''. Similarly, Christopher Blum, as chief investment officer of the U.S. Behavioral Finance Group at J.P. Morgan Fund, said that ``Humans are emotional individuals, and that gets exaggerated when it comes to taking risks\ldots These are errors that investors make, we try to exploit anomalies in valuations and momentum.'' 

The primary challenge of studying the role of investor emotions is the lack of methodology to extract direct, firm-level proxies of investor emotions. This paper is intended to provide that. We then test the connection between firm-specific investor emotions and asset price movements. Specifically, we explore whether firm-specific investor emotions predict daily price movements. Due to data and methodological difficulties measuring investor emotions, studies mainly relied on indirect proxies or were limited to experimental evidence. By pairing a large, novel dataset with recent advances in text processing, we are able to overcome the challenges inherent in studying investor emotions. In particular, we use data from StockTwits, a social networking platform for investors to share stock opinions. A critical feature of this data is that it contains firm-specific messages to compute firm-specific emotions. We employ a broad sample of over 88 million messages that span 2010 - 2021 September.

Our contribution is that we develop a simple tool that quantifies investor emotions from financial social media text data (i.e., informal text containing less than 64 words). To do so, we annotate ten thousand short messages from StockTwits and combine them with open-source emotion data. Our models are powered by deep learning and a large, novel dataset of investor messages.\footnote{For other applications of deep learning in economics see \cite{albanesi2019predicting}, \cite{vamossyemotions}, and \cite{lavoice2019racial}.} In particular, our tool takes social media text as inputs, and for each message, it constructs emotion variables corresponding to seven emotional states: neutral, happy, sad, anger, disgust, surprise, and fear.\footnote{The emotions in this paper correspond to the seven emotional states specified in \cite{breaban2018emotional}.} Our emotion variables are probabilistic measures; hence, the seven emotions sum up to 1. As an illustration, if the user supplies the text ``not feeling it :)", our model returns a tuple of (0.064, 0.305, 0.431, 0.048, 0.03, 0.038, 0.084), corresponding to (neutral, happy, sad, anger, disgust, surprise, fear). Swapping the emoticon from :) to :(, and hence the input to "not feeling it :(", our model output changes to (0.05, 0.015, 0.843, 0.009, 0.015, 0.006, 0.062). Notice that sad increased from 43.1\% to 84.3\%, while happy decreased from 30.5\% to 1.5\%. This illustrates how our model incorporates emoticons in extracting emotions from text. For additional examples, visit an \href{https://huggingface.co/vamossyd/bert-base-uncased-emotion}{interactive interface} of our model.

Using our tool, we can test whether firm-specific investor emotions predict a firm's daily price movements before the market opens. To do so, we first use our tool to generate emotion variables by quantifying the content of each message using textual analysis and then average the textual analysis results across all messages by firm-day-market open/close. Once the emotion variables are constructed, we define valence as the difference between the sum of positive (i.e., happy) and negative (i.e., sad, anger, disgust, fear) emotions. We use this metric instead for most of our analysis to ease interpretability. We then use a fixed effects model, exploiting within-firm variation in investor emotions, to test whether emotions predict a firm's daily price movements. Using within-firm variation shuts down some mechanisms. For instance, if a firm tends to have high returns, investors might always be more excited before the market opens, but including fixed effect means we can rule out that this drives our results. We also control for calendar day fixed effects to reject that our results are only driven by factors that effect emotions and returns across all firms simultaneously. We take several steps to mitigate additional estimation concerns. For instance, we rule out reactive emotions by looking at the impact of emotions before the market opens (those posted between 4:00 pm the previous trading day (i.e., $t-1$) and 9:29 am the day of (i.e., $t$) on daily trading behavior so that we have a clear temporal separation between the independent and dependent variables. We also tackle misattribution - the concern that our emotion measures are not capturing emotions correctly by investigating the relationship between our emotion measure and self-tagged bullish/bearish texts, and by investigating the impacts of contemporaneous emotions and asset prices. We find that our algorithm classifies messages happier for assets that have gone up in value.

Our analysis focuses on the role of investor emotions in explaining daily asset price movements. We document two main findings. First, we document that within-firm investor emotions can predict the company's daily price movements. For instance, one of our findings is that variation in investor enthusiasm is linked with marginally higher daily returns. Second, we find emotions behave similarly to sentiment; the impacts are larger when volatility or short interest are higher.

%% file: Sections/Related.tex
\section{Related Literature} \label{sec:lit}

\cite{ekman1992argument} defined a model of basic emotions and described the primary constituents as being: anger, disgust, fear, happiness, sadness, and surprise. This model of basic emotions has been influential in facial and text-based emotion detection (TBED).\cite{zad2021emotion} offers an insightful overview of recent progress in TBED, revealing eighteen accessible emotion-related datasets, but notably, none incorporate financial social media text. As such, our contribution is a comprehensive annotation of ten thousand financial social media posts made available for public use.

Deep learning models are continually elevating performance benchmarks in the field of Text-Based Emotion Detection (TBED). As part of this evolving landscape, our research closely aligns with the work of \cite{hartmann2022emotionenglish}.\footnote{Similarly, \cite{chiorrini2021emotion} applied deep learning and evaluated the Bidirectional Encoder Representations from Transformers (BERT) model on Twitter data. Similarly, \cite{krommyda2021experimental} used an LSTM model and compared them with five other classifiers (i.e., SVM, XGBoost, Random Forest, Naive Bayes, Decision Tree).} They have developed a DistilRoBERTa model, fine-tuned for emotion detection, which has gained widespread use and serves as a key benchmark against which we measure our model's performance. In an effort to account for the ambiguity inherent in negative posts, we also categorize our messages as positive, neutral, or negative, enabling a more nuanced comparison. This benchmarking extends to predictions made by the FinBERT model (\cite{araci2019finbert}), which also uses the positive, neutral, negative classification system, providing a comprehensive performance evaluation. In addition, we employ chatGPT (gpt-3.5-turbo) to classify our financial social media data. The labels generated by chatGPT provides us with a secondary benchmark to evaluate our models.

Our work augments the literature on TBED in several ways. First, we offer a user-friendly open-source tool along with an associated high-performing model. Second, we contribute a new dataset annotated from financial social media for use by other researchers, thereby enriching the TBED research resources. Third, we show our model outperforms competing models in detecting emotions on financial social media data.

Our research contributes to the behavioral finance literature by exploring the link between market movements and emotional states. Unlike previous studies that inferred emotions from indirect proxies such as seasonal changes (\cite{kamstra2003winter}, \cite{jacobsen2008weather}), our analysis draws from social media posts directly related to specific stocks, offering a more accurate emotional measure tied to the firm. Previous studies using similar direct proxies have generally examined the correlation between market-level investor sentiment and daily index returns (\cite{bollen2011twitter}, \cite{gilbert2010widespread}) . We, however, go a step further, demonstrating that company-specific investor emotions can forecast the firm's daily price movements. Our work also aligns with findings from laboratory experiments that investigated the role of emotions in forming market bubbles (\cite{breaban2018emotional}, \cite{andrade2016bubbling}). We provide evidence that emotions expressed in investor social media exchanges echo those observed in the controlled settings of a lab. This work, therefore, not only adds depth to the understanding of how emotions drive market behavior but also bridges the gap between observational studies and experimental settings in financial research.

The remainder of the paper is organized as follows. Section \ref{sec:text} outlines our text analysis methodology, Section \ref{sec:primary} presents our applications, and Section \ref{sec:conclusion} concludes.

%% file: Sections/Validation.tex
\section{Methodology}\label{sec:text}

We now briefly describe the text analysis methodology used in this paper. For additional details, see Appendix \ref{app:NLP}.

\subsection{Model}
Transfer learning using large-scale language models, like those based on the Transformer architecture (\cite{vaswani2017attention}), has significantly improved the performance of most NLP tasks. The Transformer architecture replaces recurrent layers with multi-headed self-attention, allowing more efficient and accelerated model training. The BERT model (\cite{devlin2018bert}) adopted this architecture to pre-train deep bidirectional representations from unlabeled text. Despite its efficiency, BERT's size can be a limitation. Therefore, we adopt DistilBERT (\cite{sanh2019distilbert}), a smaller and more cost-effective version of BERT, with one key modification - we include 4,861 tokens not present in BERT's original 30,522-token embedding.

DistilBERT retains 95\% of BERT's performance despite having half its parameters. Its efficiency is attributed to knowledge distillation, a technique where a smaller model is trained to mimic a larger model's behavior (\cite{bucilua2006model}, \cite{hinton2015distilling}).

\subsection{Model Training Data}\label{online_app:data_sources}

We first use an emotion meta-data, compiled by \cite{Bostan2018}, comprising 250K short messages. This data was created by aggregating several emotion datasets with different annotation schemes. Additionally, we use a dataset of 10,000 hand-tagged StockTwits posts, which we annotated and have made publicly accessible. In addition to human annotation, we also leveraged the capabilities of chatGPT (specifically, gpt-3.5-turbo) to classify our StockTwits training data. This offers an additional benchmark against which we can assess the performance of our models.\footnote{Details about the prompts used for gpt-3.5-turbo can be found in Appendix \ref{app:chatGPT}.}
 
We use these datasets, apply a categorical-cross-entropy loss function, and fine-tune a pre-trained DistilBERT model on human annotated labels to obtain a probabilistic assessment for each message in the data. Since our StockTwits data is small, we use the model trained on the emotion meta-data and transfer it to the StockTwits data. 

\subsection{Model Comparison}\label{app:comparison}

We now compare our emotion model, EmTract, with alternative ML models, including DistilRoBERTa fine-tuned for emotion detection (\cite{hartmann2022emotionenglish}), BERT, Random Forest (RF), Gradient Boosted Trees (XGB), Decision Trees (CART)\footnote{For RF, XGB, and CART take a few additional steps to process the text. In particular, we remove stop words, apply a Porter Stemmer, and use a TfidfVectorizer with N-grams up to 3.}.

Table \ref{tab:classification_test} reports the results. Although we train our models on human annotated data, we also evaluate the performance on chatGPT annotated labels. Panel A uses Ekman's basic emotions for performance comparison. Using five-fold cross-validation on 10,000 hand-tagged StockTwits messages, EmTract outperformed all models, with the lowest loss (1.009) and highest accuracy (64.53\%), demonstrating effective emotion detection. BERT showed similar performance, with a slight underperformance. The key distinction between these two models is that the EmTract model incorporates emojis and emoticons into the embeddings, unlike the BERT model. This inclusion plays a critical role in the EmTract model's performance, underscoring the importance emojis and emoticons play in enhancing emotion detection accuracy. Therefore, this not only validates the effectiveness of the EmTract model but also illuminates the substantial impact of visual elements like emojis and emoticons in emotion recognition tasks. \cite{hartmann2022emotionenglish}'s model, not trained on financial data, performed less effectively on StockTwits data, indicating EmTract's better handling of such nuances. Traditional ML models, RF, XGB, and CART, performed less well, with Decision Trees showing the poorest results. 

To account for the ambiguity inherent in negative emotions, we also categorize our messages as positive (happy), neutral (surprise, neutral), or negative (sad, anger, fear, disgust), enabling a more nuanced comparison. Given these classes, we can benchmark against the FinBERT model (\cite{araci2019finbert}), which also uses the positive, neutral, negative classification system. These findings are detailed in Panel B of Table \ref{tab:classification_test}. It is clear from the results that our model, even when limited to these three classes, surpasses competing models in terms of performance on both human and chatGPT annotated data.

Given that our models are tailored to human-annotated data, and that EmTract demonstrates impressive performance on such data, any divergence between the labels produced by chatGPT and human annotators could potentially impede EmTract's effectiveness when tested on chatGPT annotated data. In fact, we observe a narrowing performance gap when the model is evaluated on data labeled by chatGPT. Interestingly, \cite{hartmann2022emotionenglish} and \cite{araci2019finbert} perform similarly on human and chatGPT annotated data.

\input{Tables/five_fold}

\subsection{Model Performance}

A commonly used performance metric in the machine learning and statistics literature is an $n \times n$ contingency table, often referred to as the \emph{confusion matrix}, that describes the statistical behavior of any classification algorithm. In our application, the rows correspond to the actual class label, while the columns correspond to classifications made by our model. If a predictive model is applied to a set of messages, each message falls into one of the cells in the confusion matrix; thus, the model's performance can be assessed by the relative frequencies of the entries. 

\begin{figure}[htbp] 

\begin{center}
\includegraphics[scale=0.45]{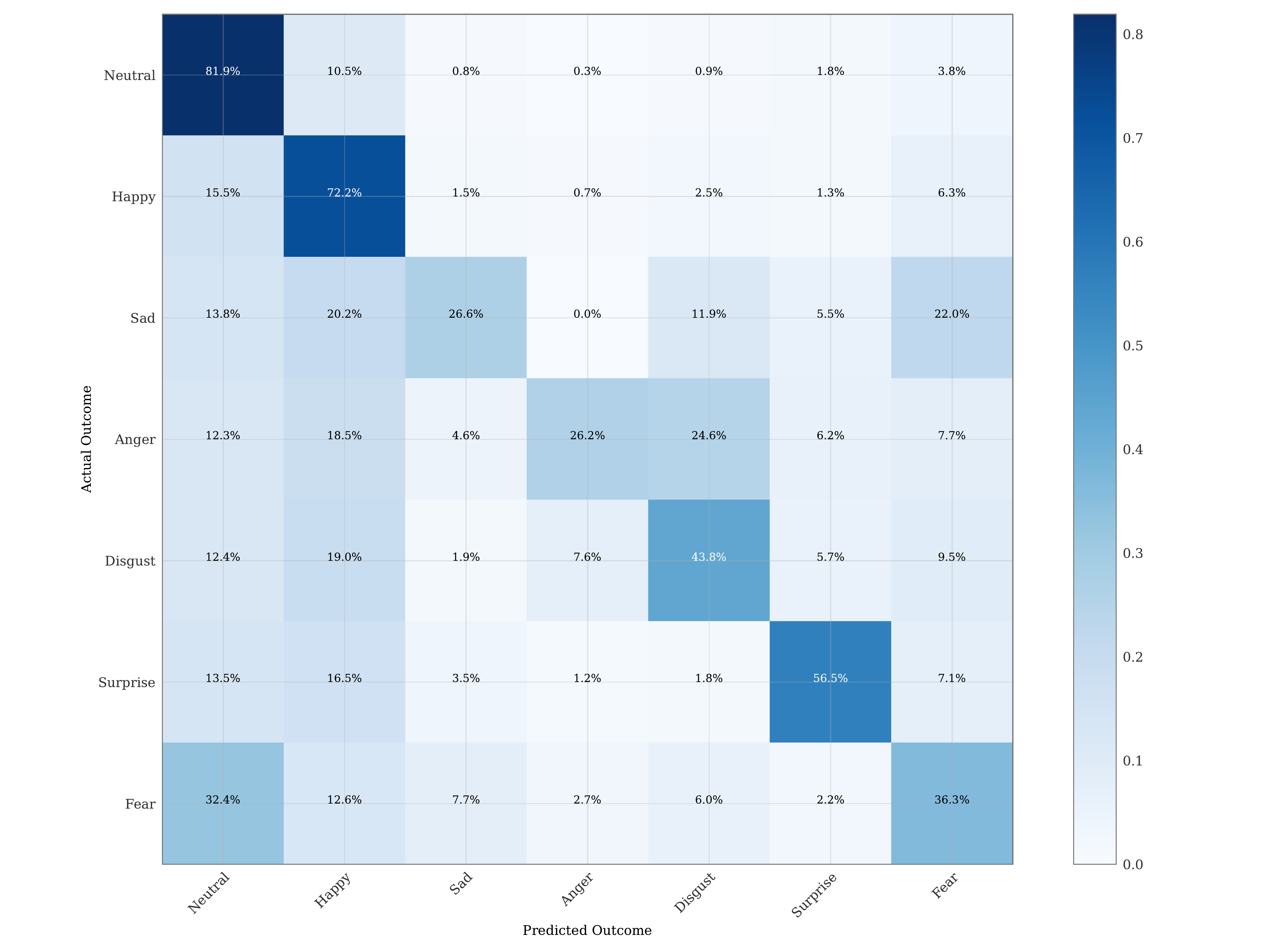}

\caption{Confusion Matrix for Emotion Classification Model (EmTract).}\label{fig:confusion_matrix}
\end{center}
  \begin{flushleft}
	\footnotesize{Notes: Results reported are based on performance on the hand-tagged sample for the best-performing model on the validation set during five-fold cross-validation.}
  \end{flushleft}
\end{figure}

As an illustration,  Figure \ref{fig:confusion_matrix} plots the confusion matrix for our StockTwits-based model. We normalize each row to sum up to 1, and hence the diagonal entries represent the classifier's precision. For instance, the 81.8\% in the upper left corner implies that our classifier accurately classified 81.9\% of all neutral messages as neutral. At the same time, the 15.5\% in the second-row first column represents that our classifier mistakenly tagged 15.5\% of happy messages as neutral. 

\subsection{Validation}
We now validate our StockTwits-based emotion model using the shut-down of meme-stock trading on January 28th, 2021. We included this exercise to show that our NLP model is capturing emotions in a way that is intuitive. We quote Matt Levine of Bloomberg to set up the background:
\begin{displayquote}
Yesterday many big retail brokerage firms told their customers that they would no longer be able to buy GameStop Corp. stock because it was getting too crazy. This led to a lot of outrage from people who are famous and online:

\begin{displayquote}
Among others rebuking the moves by the brokerages were Rep. Alexandria Ocasio-Cortez (D., N.Y.), Sen. Ted Cruz (R., Texas), billionaire Mark Cuban and Dave Portnoy, the founder of the popular digital media company Barstool Sports Inc. Mr. Portnoy was one of countless individual investors who dove into the markets this year, often streaming his trades to his followers on Twitter.

Ms. Ocasio-Cortez tweeted: “We now need to know more about @RobinhoodApp’s decision to block retail investors from purchasing stock while hedge funds are freely able to trade the stock as they see fit.”
\end{displayquote}

The popular story here is that small-time retail investors, who stereotypically trade stock on the Robinhood app and egg each other on in the WallStreetBets Reddit forum, have been pushing up the stock of GameStop for a few weeks, and they got it to pretty dizzying heights: GameStop started the year at \$18.84 per share; at one point yesterday morning it traded at \$483, up almost 2,500\%. They are doing this partly for fun and partly for profit but also, especially, to mess with the hedge funds on the other side of the trade, who had bet against GameStop by shorting the stock and who suffered and surrendered as it went up \ldots 
\end{displayquote}

\begin{figure}
\begin{center}
	\begin{subfigure}[b]{0.9\textwidth}
		\includegraphics[scale=0.35]{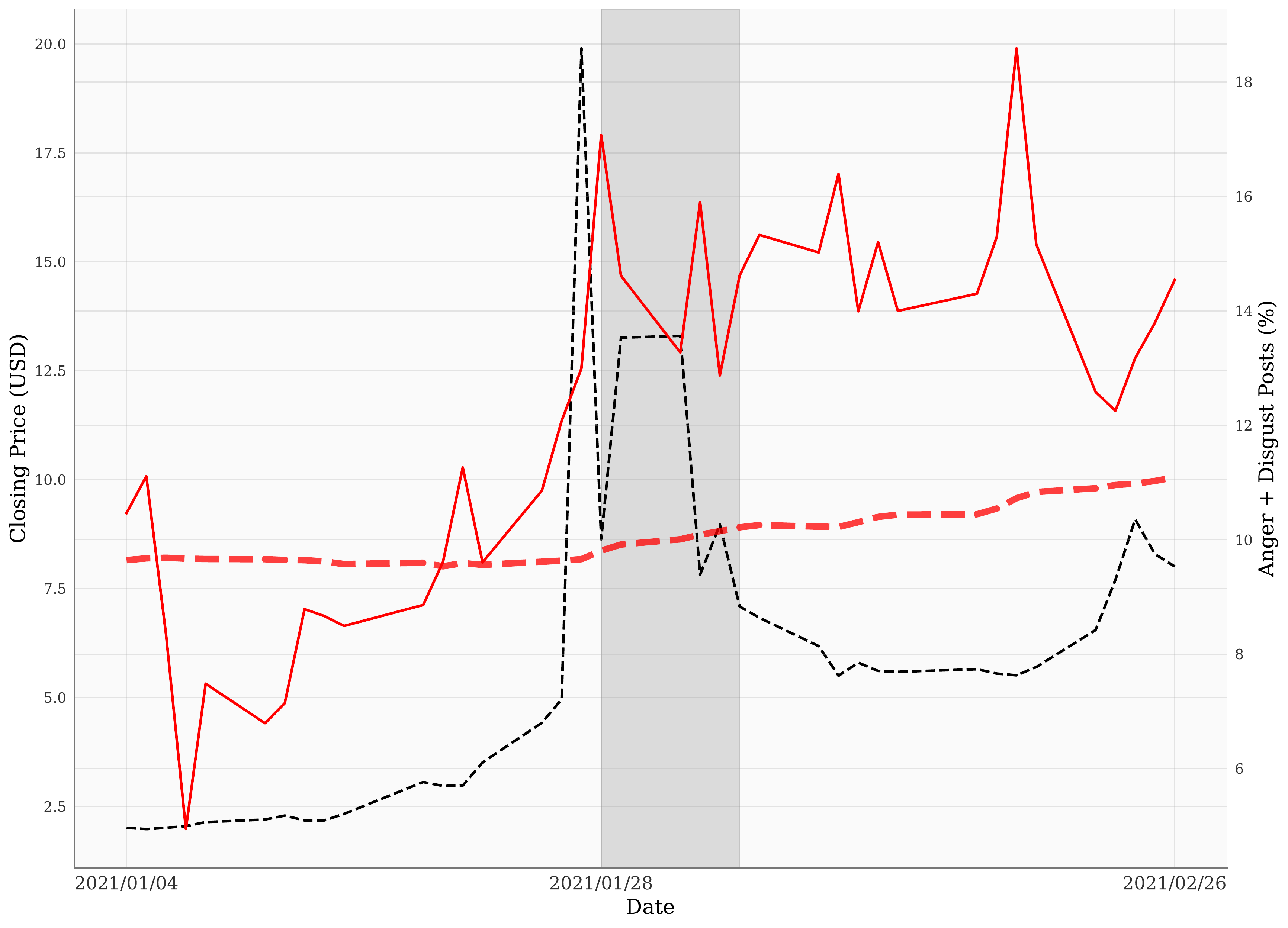}
		\caption{AMC: Anger and Disgust}
	\end{subfigure}
	
	\begin{subfigure}[b]{0.9\textwidth}
		\includegraphics[scale=0.35]{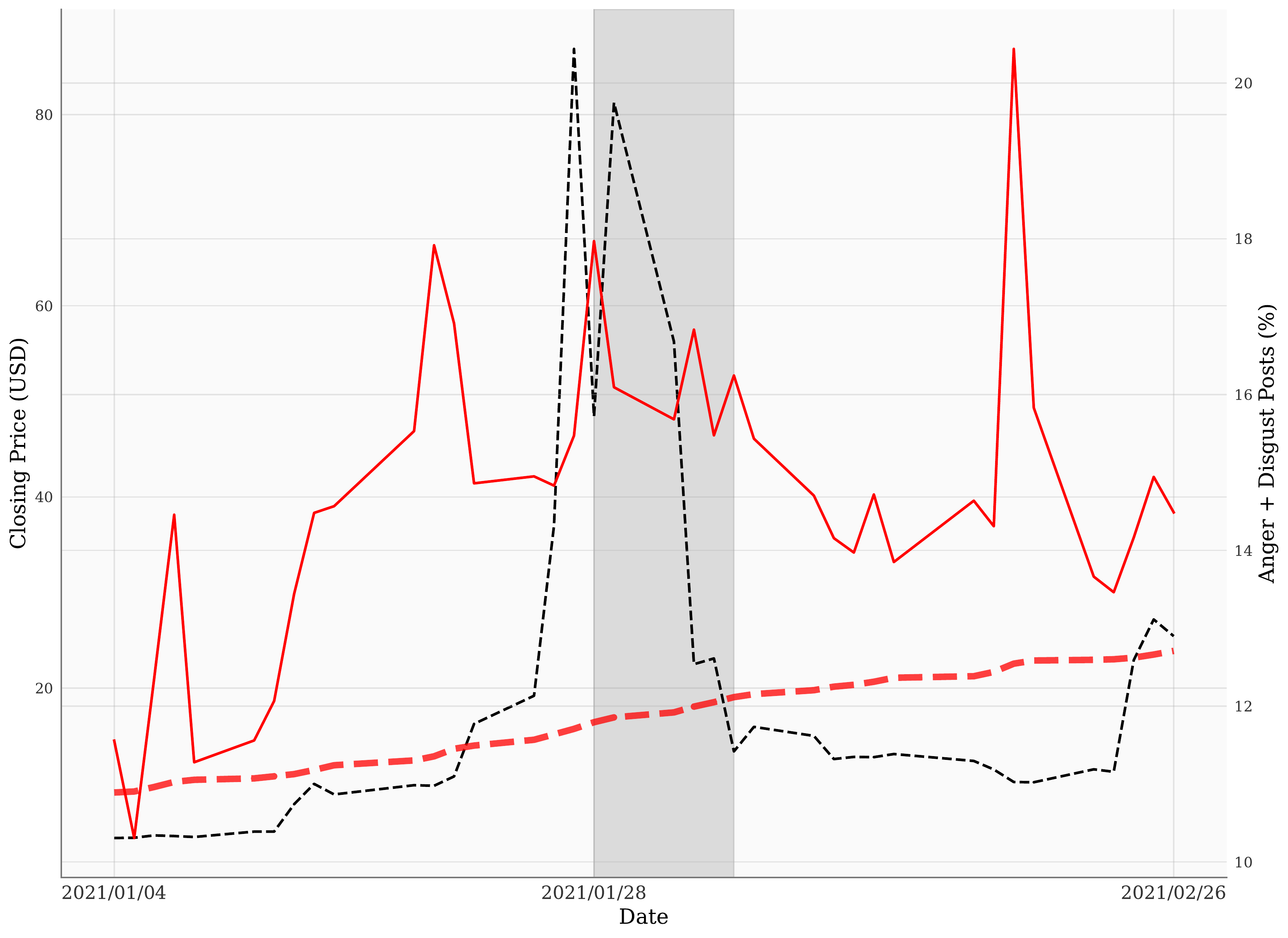}
		\caption{GME: Anger and Disgust}
	\end{subfigure}
	\end{center}
	
	\caption{Firm-Specific Anger and Disgust for Selected Meme Stocks}\label{fig:validation}
		\begin{flushleft}
		 \footnotesize{Notes: Relationship between the daily closing price and firm-specific emotions during market hours. Grey area corresponds to the dates when Robinhood restricted trading for a number of ``meme'' stocks. Dotted lines correspond to closing prices; solid lines correspond to emotions; dotted solid lines correspond to the 90 trading day average emotion. These patterns are comparable for other ``meme'' stocks. For a full list of stock restrictions, click \href{https://www.cnbc.com/2021/01/29/stocks-on-the-robinhood-restricted-trading-list-like-amc-and-koss-are-surging.html}{here}. Source: Authors' calculations based on StockTwits \& CRSP data.}
		\end{flushleft}
\end{figure}

Given the circumstances surrounding the event, we anticipate a surge in negative emotions, particularly disgust and anger, expressed on social media regarding ``meme" stocks on January 28th, 2021, and the subsequent days. In Figure \ref{fig:validation}, we present a visualization depicting the daily closing prices and selected emotions for two prominent meme stocks, namely AMC and GME, spanning from January 4th, 2021, to March 1st, 2021. The black dotted lines indicate the daily closing prices, while the colored lines represent the aggregated daily emotions. The thick colored dotted lines denote the 90-day average for the respective emotion.

As the trading restrictions were implemented, we observed a significant increase in posts containing angry and disgusted content. For instance, on January 28th, the proportion of daily posts containing angry content for AMC surged from 4\% to slightly over 9\%, which corresponds to an increase of approximately 2.4 standard deviations. Similarly, the combined proportion of anger and disgust escalated from 8-9\% to nearly 17\% for AMC. Even after the restrictions were lifted, the levels of anger and disgust gradually declined but remained elevated. The sentiment of ``the system is rigged" reverberated for a considerable period following the trading shutdown.

%% file: Tables/five_fold.tex
\begin{table}[htbp]\centering
\footnotesize
\begin{threeparttable}
\caption{Performance on Test Sample}\label{tab:classification_test}
\begin{tabular}{lllll}
 & \multicolumn{2}{c}{Human Labeled}  & \multicolumn{2}{c}{GPT 3.5 Labeled} \\
Model & Loss & Accuracy & Loss & Accuracy \\ \hline 
& & & &   \\[\dimexpr-\normalbaselineskip+2pt]
\underline{Panel A: Ekman's Emotions} & & & & \\
EmTract & 1.009 [0.059] & 64.53\% [1.98\%] & 1.463 [0.032] & 51.38\% [1.65\%] \\
BERT &  1.112 [0.029] & 61.12\% [1.65\%] & 1.453 [0.045] & 48.50\% [0.59\%]\\ 
\cite{hartmann2022emotionenglish} & 2.015 [0.047] & 45.19\% [0.94\%] & 1.935 [0.029] & 46.49\% [0.43\%]\\ 
RF &       1.421 [0.017] & 50.30\% [1.3\%]  & 1.485 [0.021] & 47.00\% [0.9\%]\\ 
XGB &        1.367 [0.022] & 52.40\% [1.59\%]  & 1.547 [0.03] & 46.17\% [1.35\%]  \\ 
CART &        18.148 [0.317] & 45.23\% [0.73\%]  & 21.32 [0.488] & 38.54\% [1.24\%] \\ \hline
& & & &   \\[\dimexpr-\normalbaselineskip+2pt]
\multicolumn{5}{l}{\underline{Panel B: Valence (Positive, Neutral, Happy)}} \\ 
& & & &   \\[\dimexpr-\normalbaselineskip+2pt]
EmTract & 0.6636 [0.028] & 72.34\% [1.44\%] & 0.932 [0.019] & 62.54\% [1.40\%] \\
\cite{hartmann2022emotionenglish} & 1.327 [0.035] & 56.70\% [0.94\%] & 1.311 [0.019] & 56.48\% [0.42\%] \\ 
\cite{araci2019finbert} & 1.995 [0.014] & 46.50\% [0.89\%] & 1.958 [0.048] & 50.24\% [0.70\%]\\ 
\hline 
\end{tabular}
\begin{tablenotes}
\footnotesize
\item Notes: The term ``Test Fold" is used to denote a subset of 2,000 messages that the model has not been exposed to during its training process. It is important to note that our models are trained using data that has been carefully labeled by human annotators. Standard deviations are represented in brackets. \cite{hartmann2022emotionenglish} and \cite{araci2019finbert} were not trained on the same dataset as ours, providing an external comparison. To ensure the robustness of our approach, we also utilized the gpt-3.5-turbo model to generate labels for our training data. Additional details regarding the prompts used for this label generation, as well as a comparative analysis between gpt-3.5-turbo generated labels and our human annotated data, can be found in Appendix \ref{app:chatGPT}.

\end{tablenotes}
\end{threeparttable}
\end{table}

%% file: Sections/Primary.tex
\section{Applications}\label{sec:primary}

As seen from Figure \ref{fig:validation}, emotions can vary considerably over time. How do such changes in emotion affect stock returns? This is the question that we examine in this section. We expect investor enthusiasm to result in increased buying pressure to push up stock prices temporarily subsequently. We first examine such price pressure in the context of daily price movements. 

Before presenting our main results, we do not establish a causal link between social media emotions and asset price movements. Our main goal is to provide a tool that extracts investor emotions from social media data tailored for financial contexts. We use this tool for illustrative purposes. The key takeaway is that investor emotions extracted from social media contain information relevant to stock valuation not accounted for by unobservable time-invariant stock characteristics, time patterns or by recent price movements. 

\subsection{Data Sources}

\subsubsection{Social Media Data (StockTwits)}

Our investor emotion dataset comes from StockTwits, founded in 2008 as a social networking platform for investors to share stock opinions. StockTwits looks similar to Twitter, where users post messages of up to 140 characters (280 characters since late 2019), and use ``cashtags'' with the stock ticker symbol (e.g., \$AMZN) to link ideas to a particular company. Although the app does not directly integrate with other social media platforms, participants can share content to their personal Twitter, LinkedIn, and Facebook accounts.

Our original dataset's range falls between 1 January 2010 and 30 September 2021. In total, there are 242M messages. For each message, we observe sentiment indicators tagged by the user (bullish, bearish, or unclassified), sentiment score computed by StockTwits, ``cashtags'' that connect the message to particular stocks, like count, and a user identifier.\footnote{The user identifier allows us to explore characteristics of the user, such as follower count. For most users, we also have information on a self-reported investment philosophy that can vary along two dimensions: (1) Approach - technical, fundamental, momentum, value, growth, and global macro; or (2) Holding Period - day trader, swing trader, position trader, and long term investor. We group technical with momentum and value with fundamental for our heterogeneity explorations. For investment horizon, we explore short and long-term investors. Users of the platform also provide their experience level as either novice, intermediate, or professional. This user-specific information about the style, experience, and investment model employed helps explore heterogeneity in investor emotions.} We remove messages that appear automated.\footnote{We define automated messages as cleaned messages posted over 100 times by the same user over the period 2010-2021.} We focus on messages that can be directly linked to particular stocks, so we restrict attention to messages that only mention one ticker. Next, we require at least ten posts per stock per session (market and non-market) to discard noisy signals. Last, we filter out inactive tickers based on SECSTAT and restrict our analysis to common ordinary shares traded on U.S. exchanges. We summarize our sample restrictions in Table \ref{tab:sample_restrictions}. We end up with 88M messages authored by 984,434 users, covering 4,319 tickers.

\input{Tables/sample_restrictions}

\subsubsection{Short Interest, and Pricing Data}
Price-related variables are obtained from the merged CRSP/COMPUSTAT file, short interest from Compustat's Supplemental Short Interest File. We match these with StockTwits on the ticker and date. 

\subsection{Descriptive Statistics}

We follow \cite{breaban2018emotional} and define valence as:
\begin{equation}
    \text{Valence} = \text{Happy} - \text{Sad} - \text{Anger} - \text{Disgust} - \text{Fear}
\end{equation}

Table \ref{tab:summary_stats} presents the descriptive statistics for the analysis variables. 
We report the mean, the standard deviation (in parentheses), and the within-firm standard deviation [in brackets]. Panel A reports summary statistics for our social media-based variables. We find a high fraction of neutral messages, with a mean of 51.1\%, followed by 24.6\% of happy posts. Looking at the emotion variables, we observe a positive skewness. This might suggest that investors are more likely to share their enthusiasm on social media than pessimism. Fear and surprise are the third and fourth most frequent emotions, followed by disgust, sad, then anger. Similarly to our return variables, the within-firm standard deviation is slightly smaller than the sample standard deviation for our emotion variables. Hence, we can exploit plenty of variation even with a firm-date fixed effect framework. 

To help with interpreting the magnitude of our other variables, we contrast our final estimation sample with a CRSP/Compustat sample covering the same time, following the same sample restrictions (i.e., active, common ordinary shares traded on U.S. exchanges). Panel B of Table \ref{tab:summary_stats} presents the results. We observe a few interesting patterns. First, StockTwits users tend to discuss stocks that have gone up or are currently going up in value: the average past monthly return is 3.3 percentage points, the close-open return is 0.3 percentage points, the one-day lag open-close return is 0.1 percentage point higher than in the CRSP sample. Second, these stocks end up with a 0.3 percentage point lower open-close return, possibly suggesting mean-reversion. We also find that social media investors are more interested in discussing firms with higher \$ trading volume, volatility, market cap, and short interest and with lower institutional ownership and past returns. 

\input{Tables/summary_tab}

\subsection{Empirical Framework}

Our return regressions exploit within-firm variation and horse-race our valence measure with the sentiment measure computed by StockTwits. Hence, our empirical specification explores whether our valence measure provides additional information beyond sentiment.\footnote{We repeated this exercise without controlling for sentiment and replacing valence with our multi-dimensional emotion measures. Our results are qualitatively similar and remain statistically significant.} That is, we estimate the following model:
\begin{equation}\label{eq:emotion}
 Y_{it} = \alpha + \beta_0 \text{Valence}_{it} + \beta_1 \text{StockTwits Sentiment}_{it} +\gamma Y_{i,t-1} + \zeta X_{i, t-1} + \delta_t + \delta_i + \epsilon_{it} 
\end{equation}

where Valence$_{it}$ is our valence measure for firm i at date t, and StockTwits Sentiment$_{it}$ is the sentiment score provided by StockTwits.\footnote{The intercept ($\alpha$) in Equation (2) soaks up the average value of fixed effects ($\delta_i$ and $\delta_t$). Our coefficients of interest are $\beta_0$ and $\beta_1$. Controlling for both, $\beta_0$ tells us how much additional information valence (emotions extracted by our model) give us beyond the sentiment score provided by StockTwits.} Our text-based measures are computed based on messages during non-trading hours before the market opens. Our dependent variable for most of our analysis is the open-close daily return. Our day ($\delta_t$) fixed effects to control for important daily news events (or for the daily market return) at the market level, while firm ($\delta_i$) fixed effects shut down some mechanisms. For instance, if a firm tends to have high returns, investors might always be more excited before the market opens, but including firm fixed effects mean we can rule out that this drives our results. We account for autocorrelation by controlling for the open-close return on day $t-1$. We include the close-open return (i.e., the previous day's close to the current open) to ensure that our emotions contain incremental information to after-hours and pre-market price movements. Most of our specifications also include controls ($X_{i, t-1}$) for the past 20 trading days' return and volatility as defined in Table \ref{tab:variable_definitions}. To mitigate the concern that our emotion measure only reflects changes in trading activity, we regress our dependent variable on day t (i.e., open-close return) on emotion among messages posted before the market opened on day t. In this case, the emotion measure leads the dependent variable in time.

\subsection{Emotions, Information Content, and Returns}

We now explore the relationship between investor emotions and daily stock returns with our regression framework. Closest to our setting, lab evidence finds a positive (negative) relationship between pre-market happiness (fear) and returns (\cite{breaban2018emotional}). We evaluate this empirically in Table \ref{tab:pre_market}. Column (1) includes our control variables, sentiment, firm, and date fixed effects. The direct comparison between Columns (1) and (2) shows that emotions before the market opens can explain a small fraction of the variation in daily returns ($R^2$ increases by 0.02 percentage points). In addition, a standard deviation increase in non-market hour valence (before the market opens on day t) is associated with a 1.5\% standard deviation increase in daily stock returns. Column (3) shows that the point estimates of investor emotions and sentiment are smaller for larger and easier-to-value firms, i.e., S\%P Super Composite firms. Last, we find stronger effects for stocks with larger user engagement (Column (4)), i.e., at least 100 messages, such that a standard deviation increase in happiness before the market opens on day t is associated with a 3.3\% standard deviation increase in daily open-close stock returns. These findings suggest that investor emotions extracted from StockTwits provide information relevant to stock valuation not accounted for by unobservable time-invariant stock characteristics, time patterns, recent price movements, or by sentiment. 

\input{Tables/pre_market_ret}

Table \ref{tab:pre_market_content} repeats the analysis using measures of emotions disaggregated between messages that convey earnings or trade-related information (``finance'') and messages that provide other information (``chat'') (Column (1) and Column (2)). We find that emotions extracted from messages specifically mentioning firm fundamentals and earnings typically have smaller point estimates. This suggests that important information is conveyed in seemingly irrelevant messages we capture by extracting emotions. For instance, a user may post ":) :)" as a reaction to news events or other posts, which clearly expresses her excitement yet do not contain any information about the stock aside from her emotion. This is in stark contrast with the StockTwits provided sentiment, for which posts containing financial language have a five times larger point estimate. Next, contrasting Column (3) and Column (4), we find that both messages containing original information and those disseminating existing information drive our results. Interestingly, however, the relationship by information type also suggests differences between our valence measure and the StockTwits-provided sentiment measure. In particular, for sentiment it is disseminating existing information, while for valence it is original content that has larger point estimate. 

\input{Tables/pre_market_content_ret}

\subsection{Heterogeneous Effects}
Thus far, the results suggest that social media investor emotions provide valuable
information that can help explain daily price movements. However, this effect is unlikely to be uniform. In this section, we explore heterogeneous effects concerning stock characteristics.  Theory on investor sentiment posits that younger, smaller, more volatile, unprofitable, non–dividend paying, distressed stocks are most sensitive to investor sentiment. Conversely, ``bond-like'' stocks are less driven by sentiment (see, \cite{baker2007investor}). 

To examine whether emotions behave similarly to sentiment, we first interact our emotion variables with a dummy variable intended to capture high (above the median) volatility and low (below the median) liquidity (i.e., market cap). In line with the theory on investor sentiment, we find larger point estimates for more volatile (Column (1-2)) and lower market cap firms (Column (3-4) of Table \ref{tab:pre_market_volat}). When interacting our key explanatory variables with an indicator variable of being over the median in short interest, we find larger impacts when short interest is higher (Column (5-6) of Table \ref{tab:pre_market_volat}). 

\input{Tables/pre_market_volatility_ret}

%% file: Tables/sample_restrictions.tex
\begin{table}[htbp]\centering
\def\sym#1{\ifmmode^{#1}\else\(^{#1}\)\fi}
\begin{threeparttable}
\footnotesize
\caption{Itemized Sample Restrictions}\label{tab:sample_restrictions}
\begin{tabular}{lll}
 &  & Messages \\ \hline 
   &  &  \\
StockTwits Data 2010-2021* &  & 242,362,477
 \\ \hline
     &  &  \\
 Keep & &  \\ \hline 
   &  &  \\
Single Ticker
 &  & 180,298,172
 \\
Not Automated
&  & 167,040,745
 \\ 
Active, Common Ordinary Shares, Traded on US Exchanges**
&  & 103,248,233
 \\ 
At Least 10 Messages during Non-Market/Market Hours
&  & 88,055,174

 \\ \hline 
   &  &  \\
Final Pre-Market (4pm-9am) Sample & & 37,657,214          \\ 
   & & \\
Final Market (9am-4pm) Sample & & 50,397,960 \\ 
   & & \\ \hline \hline 
\end{tabular}
\begin{tablenotes}
\footnotesize 
\item Notes: \sym{*}Sample starts January 1st 2010 and runs till September 30th 2021. \sym{**}We filter out stocks for which the security status (SECSTAT is ``I") is inactive and restrict our sample to common ordinary shares (TPCI is ``0") traded on US exchanges (EXCHG is 11, 12, 14, or 17).
\end{tablenotes}
\end{threeparttable}
\end{table}

%% file: Tables/summary_tab.tex
\begin{table}[htbp]\centering
\begin{threeparttable}
\footnotesize 
\def\sym#1{\ifmmode^{#1}\else\(^{#1}\)\fi}
\caption{Summary Statistics\label{tab:summary_stats}}
\begin{tabular}{l*{3}{c}}
\hline\hline
& & &\\[\dimexpr-\normalbaselineskip+2pt]
                    &\multicolumn{1}{c}{CRSP/Compustat-StockTwits}&\multicolumn{1}{c}{CRSP/Compustat}&\multicolumn{1}{c}{Difference} \\
\hline
& & &\\[\dimexpr-\normalbaselineskip+2pt]
\multicolumn{4}{l}{\textbf{Panel A: Social Media Information}} \\
& & &\\[\dimexpr-\normalbaselineskip+2pt]
\textcolor{white}{...}Happy               &       0.246& &\\
                    &     (0.150) [0.108] & &\\
& & &\\[\dimexpr-\normalbaselineskip+2pt]
\textcolor{white}{...}Sad                 &         0.036&     &   \\
                    & (0.0294) [0.0257]  &   &\\
& & &\\[\dimexpr-\normalbaselineskip+2pt]
\textcolor{white}{...}Fear                &        0.085&   &    \\
                    & (0.0532) [0.0432]   &   &\\
& & &\\[\dimexpr-\normalbaselineskip+2pt]
\textcolor{white}{...}Disgust             &        0.037&    &   \\
                    &  (0.0343) [0.0284]  &   &\\
& & &\\[\dimexpr-\normalbaselineskip+2pt]
\textcolor{white}{...}Anger               &        0.024&    &   \\
                    &  (0.0213) [0.0174] &   &\\
& & &\\[\dimexpr-\normalbaselineskip+2pt]
\textcolor{white}{...}Surprise            &        0.062&   &    \\
                    &  (0.0511) [0.0438]  &   &\\
& & &\\[\dimexpr-\normalbaselineskip+2pt]
\textcolor{white}{...}Neutral             &        0.511&   &    \\
                    &   (0.248) [0.155]  &   &\\
\textcolor{white}{...}Valence             &         0.065&   &    \\
                    &  (0.150) [0.137]  &   &\\
\textcolor{white}{...}StockTwits Sentiment             &         0.078&   &    \\
                    &    (0.200) [0.187]  &   &\\
                    
& & &\\[\dimexpr-\normalbaselineskip+2pt]
\multicolumn{4}{l}{\textbf{Panel B: Financial Information}} \\
& & &\\[\dimexpr-\normalbaselineskip+2pt]
\textcolor{white}{...}Open-Close Return  & -0.002    & 0.000   & -0.003\sym{***}  \\
                    &  (0.0508) [0.0481] & (0.032)  &\\
& & &\\[\dimexpr-\normalbaselineskip+2pt]
\textcolor{white}{...}Close-Open Return &     0.004   &  0.001    &      0.003\sym{***}\\
            &   (0.0410) [0.0390]  &  (0.022)  &                     \\
& & &\\[\dimexpr-\normalbaselineskip+2pt]
\textcolor{white}{...}Open-Close Return$_{-1}$      &   0.001    &     0.000  &      0.001\sym{***}\\
            &   (0.0536) [0.0507] &   (0.032)  &                     \\
& & &\\[\dimexpr-\normalbaselineskip+2pt]
\textcolor{white}{...}Return$_{-20,-1}$        &       0.047 &      0.014&           0.033\sym{***}   \\
                    &  (0.297) [0.267] & (0.158)   &\\
& & &\\[\dimexpr-\normalbaselineskip+2pt]
\textcolor{white}{...}\$ Volume$_{-183,-1}$ &     17.032&   15.229 &   1.803\sym{***}   \\
                    &   (2.615) [0.986]  & (2.764)  &\\
& & &\\[\dimexpr-\normalbaselineskip+2pt]
\textcolor{white}{...}Volatility$_{-183,-1}$    &    0.039&   0.027&      0.012\sym{***}      \\
                    & (0.0217 [0.00932]  &  (0.017)  &\\
& & &\\[\dimexpr-\normalbaselineskip+2pt]
\textcolor{white}{...}Market Cap$_{-1}$     &       21.398&    20.742&      0.655\sym{***}      \\
                    &  (2.624) [0.921]  & (2.076) &\\
& & &\\[\dimexpr-\normalbaselineskip+2pt]
\textcolor{white}{...}Institutional Ownership&          0.518&   0.602&     -0.084\sym{***}   \\
                    &   (0.322) [0.106] & (0.318) &\\
& & &\\[\dimexpr-\normalbaselineskip+2pt]
\textcolor{white}{...}Short Interest    &         0.074&    0.042&   0.032\sym{***}    \\
                    & (0.0819) [0.0478]  &   (0.053) &\\
\hline
 Observations       & 479,463 &       10,428,859  &            \\
\hline\hline
\end{tabular}
   \begin{tablenotes}
     \footnotesize
    \item Notes: \$ Volume and Market capitalization are in logs. Institutional Ownership, Short Interest, Return, and Volatility are in units. Close-Open Return refers to the return between closing price at $t-1$ to the opening price at $t$. Valence is defined as the sum of negative emotions (i.e., sad, anger, disgust, fear) subtracted from the positive ones (i.e., happy). StockTwits Sentiment is provided from StockTwits, -1 indicates bearich, while 1 indicates bullish. \sym{*} \(p<0.10\), \sym{**} \(p<0.05\), \sym{***} \(p<0.01\). Standard deviations in parentheses, within-firm standard deviation in brackets for our CRSP/Compustat-StockTwits sample only (we remove firm and date fixed effects to help interpret effect sizes). 
   \end{tablenotes}
\end{threeparttable}
\end{table}

%% file: Tables/pre_market_ret.tex
\begin{table}[htbp]\centering
\footnotesize
\begin{threeparttable}
\def\sym#1{\ifmmode^{#1}\else\(^{#1}\)\fi}
\caption{Pre-Market Emotions and Price Movements \label{tab:pre_market}}
\begin{tabular}[lcccc]{p{1.5in}p{1in}p{1in}p{1in}p{1in}}

\hline\hline
 &                     &                     &                                    \\[\dimexpr-\normalbaselineskip+2pt]
                    &\multicolumn{1}{c}{(1)}&\multicolumn{1}{c}{(2)}&\multicolumn{1}{c}{(3)}&\multicolumn{1}{c}{(4)}\\
\hline
                     &                     &                     &                    &                       \\[\dimexpr-\normalbaselineskip+2pt]
StockTwits Sentiment     &      0.0065\sym{***}&      0.0055\sym{***}&      0.0019\sym{***}&      0.0211\sym{***}\\
                    &    (0.0003)         &    (0.0003)         &    (0.0004)         &    (0.0048)         \\
                     &                     &                     &                    &                       \\[\dimexpr-\normalbaselineskip+2pt]
EmTract Valence             &                     &      0.0051\sym{***}&      0.0021\sym{***}&      0.0161\sym{***}\\
                    &                     &    (0.0006)         &    (0.0007)         &    (0.0043)         \\
\hline
                     &                     &                     &                    &                       \\[\dimexpr-\normalbaselineskip+2pt]
S\&P Super Composite Firms      &                     &                     &           X         &                     \\
At least 100 messages&                     &                     &                     &           X         \\ \hline 
                     &                     &                     &                    &                       \\[\dimexpr-\normalbaselineskip+2pt]
$\sigma_{y, within}$&         0.0475         &      0.0475         &      0.0308         &      0.0669         \\
Observations        & 454138        & 454138        & 163006        &  53803        \\
$R^2$               &      0.1053         &      0.1055         &      0.1443         &      0.1743         \\
\hline\hline
\end{tabular}
\begin{tablenotes}
\footnotesize 
\item Notes: This table considers the relationship between valence, sentiment and daily price movements. The dependent variable is daily returns, computed as the difference between the closing and the opening price, divided by the opening price. All specifications include firm and date fixed effects, close-open returns, lag open-close returns, past 20 trading days return and volatility (as defined in Variable Definitions). Robust standard errors clustered at the industry and date levels are in parentheses. We use the Fama-French 12-industry classification. \sym{*} \(p<0.10\), \sym{**} \(p<0.05\), \sym{***} \(p<0.01\). Continuous variables winsorized at the 0.1\% and 99.9\% level to mitigate the impact of outliers. We report the within-firm, detrended (demeaned) standard deviation of the dependent variable. 
\end{tablenotes}
\end{threeparttable}
\end{table}

%% file: Tables/pre_market_content_ret.tex
\begin{table}[htbp]\centering
\footnotesize
\begin{threeparttable}
\def\sym#1{\ifmmode^{#1}\else\(^{#1}\)\fi}
\caption{Pre-Market Emotions, Message Content and Price Movements \label{tab:pre_market_content}}
\begin{tabular}[lcccc]{p{1.5in}p{1in}p{1in}p{1in}p{1in}}
\hline\hline
 &                     &                     &                     &                \\[\dimexpr-\normalbaselineskip+2pt]
                    &\multicolumn{1}{c}{(1)}&\multicolumn{1}{c}{(2)}&\multicolumn{1}{c}{(3)}&\multicolumn{1}{c}{(4)}\\
                    & \multicolumn{2}{c}{\uline{Chat Type}} & \multicolumn{2}{c}{\uline{Information Type}}  \\ 
                    &\multicolumn{1}{c}{Chat}&\multicolumn{1}{c}{Finance}&\multicolumn{1}{c}{Disseminating}&\multicolumn{1}{c}{Original}\\
\hline
 &                     &                     &                     &                \\[\dimexpr-\normalbaselineskip+2pt]
StockTwits Sentiment     &      0.0010\sym{***}&      0.0049\sym{***}&      0.0038\sym{***}&      0.0027\sym{***}\\
                    &    (0.0003)         &    (0.0003)         &    (0.0003)         &    (0.0003)         \\
 &                     &                     &                     &                \\[\dimexpr-\normalbaselineskip+2pt]
EmTract Valence             &      0.0029\sym{***}&      0.0022\sym{***}&      0.0020\sym{***}&      0.0034\sym{***}\\
                    &    (0.0003)         &    (0.0005)         &    (0.0004)         &    (0.0003)         \\
\hline
 &                     &                     &                     &                \\[\dimexpr-\normalbaselineskip+2pt]
$\sigma_{y, within}$   &      0.0484         &      0.0475         &      0.0475         &      0.0487         \\
Observations        & 418899        & 453943        & 447756        & 413117        \\
$R^2$               &      0.1078         &      0.1054         &      0.1056         &      0.1100         \\
\hline\hline
\end{tabular}
\begin{tablenotes}
\footnotesize 
\item Notes: This table considers the relationship between pre-market emotions and sentiment disaggregated by information content and information type, and daily price movements. The dependent variable is daily returns, computed as the difference between the closing and the opening price, divided by the opening price. All specifications include firm and date fixed effects, close-open returns, lag open-close returns, past 20 trading days return and volatility (as defined in Variable Definitions). Robust standard errors clustered at the industry and date levels are in parentheses. We use the Fama-French 12-industry classification. \sym{*} \(p<0.10\), \sym{**} \(p<0.05\), \sym{***} \(p<0.01\). Continuous variables winsorized at the 0.1\% and 99.9\% level to mitigate the impact of outliers. We report the within-firm, detrended (demeaned) standard deviation of the dependent variable. 
\end{tablenotes}
\end{threeparttable}
\end{table}

%% file: Tables/pre_market_volatility_ret.tex
\begin{sidewaystable}
\centering 
\footnotesize
\begin{threeparttable}
\def\sym#1{\ifmmode^{#1}\else\(^{#1}\)\fi}
\caption{Pre-Market Emotions, Stock Characteristics and Price Movements \label{tab:pre_market_volat}}
\begin{tabular}[lccccccc]{p{2in}p{0.75in}p{0.75in}p{0.75in}p{0.75in}p{0.75in}p{0.75in}}
\hline\hline
                    &\multicolumn{1}{c}{(1)}&\multicolumn{1}{c}{(2)}&\multicolumn{1}{c}{(3)}&\multicolumn{1}{c}{(4)}&\multicolumn{1}{c}{(5)}&\multicolumn{1}{c}{(6)}\\
                    & \multicolumn{6}{c}{\uline{Interaction Variable}} \\
                    &\multicolumn{2}{c}{Volatility}&\multicolumn{2}{c}{Market Cap}&\multicolumn{2}{c}{Short Interest}\\
\hline
              &        &                     &                     &       & &         \\[\dimexpr-\normalbaselineskip+2pt]
Valence             &      0.0057\sym{***}&                     &      0.0053\sym{***}&                     &      0.0047\sym{***}&                     \\
                    &    (0.0006)         &                     &    (0.0007)         &                     &    (0.0008)         &                     \\
              &        &                     &                     &       & &         \\[\dimexpr-\normalbaselineskip+2pt]
Valence $\times$ Interaction Variable           &      0.0025\sym{**} &                     &      0.0030\sym{***}&                     &      0.0042\sym{***}&                     \\
                    &    (0.0010)         &                     &    (0.0010)         &                     &    (0.0011)         &                     \\
              &        &                     &                     &       & &         \\[\dimexpr-\normalbaselineskip+2pt]
StockTwits Sentiment       &                     &      0.0045\sym{***}&                     &      0.0035\sym{***}&                     &      0.0052\sym{***}\\
                    &                     &    (0.0003)         &                     &    (0.0003)         &                     &    (0.0004)         \\
              &        &                     &                     &       & &         \\[\dimexpr-\normalbaselineskip+2pt]
StockTwits Sentiment  $\times$ Interaction Variable &                     &      0.0055\sym{***}&                     &      0.0070\sym{***}&                     &      0.0026\sym{***}\\
                    &                     &    (0.0008)         &                     &    (0.0007)         &                     &    (0.0006)         \\
              &        &                     &                     &       & &         \\[\dimexpr-\normalbaselineskip+2pt]
Interaction Variable&       0.0002         &     -0.0000         &      0.0047\sym{***}&      0.0044\sym{***} &      0.0008\sym{***}&      0.0009\sym{***}\\
                    &    (0.0003)         &    (0.0003)         &    (0.0004)         &    (0.0004)         &    (0.0003)         &    (0.0003)         \\

\hline
 &                     &                     &                     &       & &         \\[\dimexpr-\normalbaselineskip+2pt]
$\sigma_{y, within}$&        0.0475         &      0.0475         &      0.0477         &      0.0477         &      0.0476         &      0.0476         \\
Observations        & 454328         & 454138         & 467506         & 467316         & 467196         & 467015         \\
$R^2$               &      0.1051         &      0.1054         &      0.1058         &      0.1061         &      0.1054         &      0.1056         \\
\hline\hline
\end{tabular}
\begin{tablenotes}
\footnotesize 
\item Notes: This table contrast the relationship between pre-market emotions, stock characteristics and daily price movements with pre-market sentiment, stock characteristics and daily price movements. The dependent variable is daily returns, computed as the difference between the closing and the opening price, divided by the opening price. The interaction variables are dummy variables for being above the median in terms of volatility (Column (1-2)), lower than the median in terms of market cap (Column (3-4)), and being above the median in terms of short interest (Column (5-6)). All specifications include firm and date fixed effects, close-open returns, lag open-close returns, and past 20 trading days return. Robust standard errors clustered at the industry and date levels are in parentheses. We use the Fama-French 12-industry classification. \sym{*} \(p<0.10\), \sym{**} \(p<0.05\), \sym{***} \(p<0.01\). Continuous variables winsorized at the 0.1\% and 99.9\% level to mitigate the impact of outliers. We report the within-firm, detrended (demeaned) standard deviation of the dependent variable. 
\end{tablenotes}
\end{threeparttable}
\end{sidewaystable}

%% file: Sections/Conclusion.tex
\section{Conclusion} \label{sec:conclusion}

In this paper, we develop a tool to extract investor emotions from financial social media data. The tool is open-source and is available at \href{https://github.com/dvamossy/ProjectNostradamus}{https://github.com/dvamossy/EmTract}. Using this tool, we make contributions to the literature on investor emotions and market behavior. Specifically, we find that investor emotions can help predict daily price movements. These impacts are larger when volatility or short interest are higher, and when institutional ownership and liquidity are lower. 

While the effects of information on fundamentals can be identified with well-established techniques in finance and economics, studying the emotional component requires new tools. In our view, the methods described herein constitute a step forward in this direction.
\pagebreak

%% file: Sections/References.tex
\singlespacing
\nocite{*}
\bibliographystyle{apacite}
\bibliography{bib}

\clearpage

%% file: Sections/Appendix.tex
\onehalfspacing

\appendix
\clearpage

\counterwithin{figure}{section}
\counterwithin{table}{section}
\addcontentsline{toc}{section}{Appendices}

\section*{Appendix}\label{app:app}

\section{chatGPT Annotation}\label{app:chatGPT}
In this section, we discuss how we use prompts for chatGPT3.5, and compare human with ChatGPT emotion annotations.

\subsection{Prompt for GPT 3.5}

Our approach to prompting is similar to \cite{lopez2023can}. Prompts are vital for directing ChatGPT's responses to various tasks. These short contextual texts range from single sentences to lengthy paragraphs, based on task complexity. They initiate ChatGPT's response generation process, where the model analyzes the prompt's syntax and semantics, proposes potential responses, and selects the most suitable one based on coherence, relevance, and grammatical accuracy. We set a temperature of 0 to maximize the reproducibility of the results, and apply the following prompt to classify social media posts:

\begin{lstlisting}[language=Python]
response = openai.ChatCompletion.create(
  model="gpt-3.5-turbo",
  temperature=0,
  messages=[
    {"role": "system", "content": """
You are a helpful assistant capable of analyzing and classifying emotions in social media posts, 
typically from a trading or financial context, into the following categories: neutral, happy, sad, anger, disgust, fear, and surprise. 
Note that these categories correspond to Ekman's six emotions and a neutral class. You only have these categories to choose from.

Here are some examples:
Post: $EURUSD #EURUSD screencast here: http://stks.co/c0sjB Take a FREE trial of all our #FX reports, email Sales@MarketChartist.com #Forex\n neutral\n
Post: $DRYS will they release ER pre market or after hour ?\n neutral\n
Post: $PTX so close to a breakout!!!\n happy\n
Post: $SPY...Could see a strong reversal RIGHT HERE...BE READY WARRIORS!!!!\n happy\n
Post: $HTZ no bueno anymore\n sad\n
Post: $CSCO March 27th 46$ calls are causing me much pain\n sad\n
Post: $ENSV *** you OPEC you greasy ***.\n anger\n
Post: $IZEA what a piece of actual ***. Ted can choke on one\n anger\n
Post: $NAK UH-OOOOOOOOOO. TURDS always sink BLUB BLUB\n disgust\n
Post: $HTBX get a load of this scammer *** and BLOCKED\n disgust\n
Post: $ZN WOW! Wow, & wow!!!! after hours trading, down .62 trading @ $1.45 yaaaa baby!!!!!\n surprise\n
Post: $PLUG wow, now just heartless\n surprise\n
Post: $JNUG maybe sell at the high today?\n fear\n
Post: $WTW run traders the market is going down. Fed rates tomorrow\n fear\n
    """
    },
    {"role": "user", "content": f"Here's a social media post: '{data_list[num]}'. With one word, how would you label this post in terms of emotions?"}
  ]
)
\end{lstlisting}

\subsection{Annotation Discrepancies}

\begin{figure}[htbp] 

\begin{center}
\includegraphics[scale=0.45]{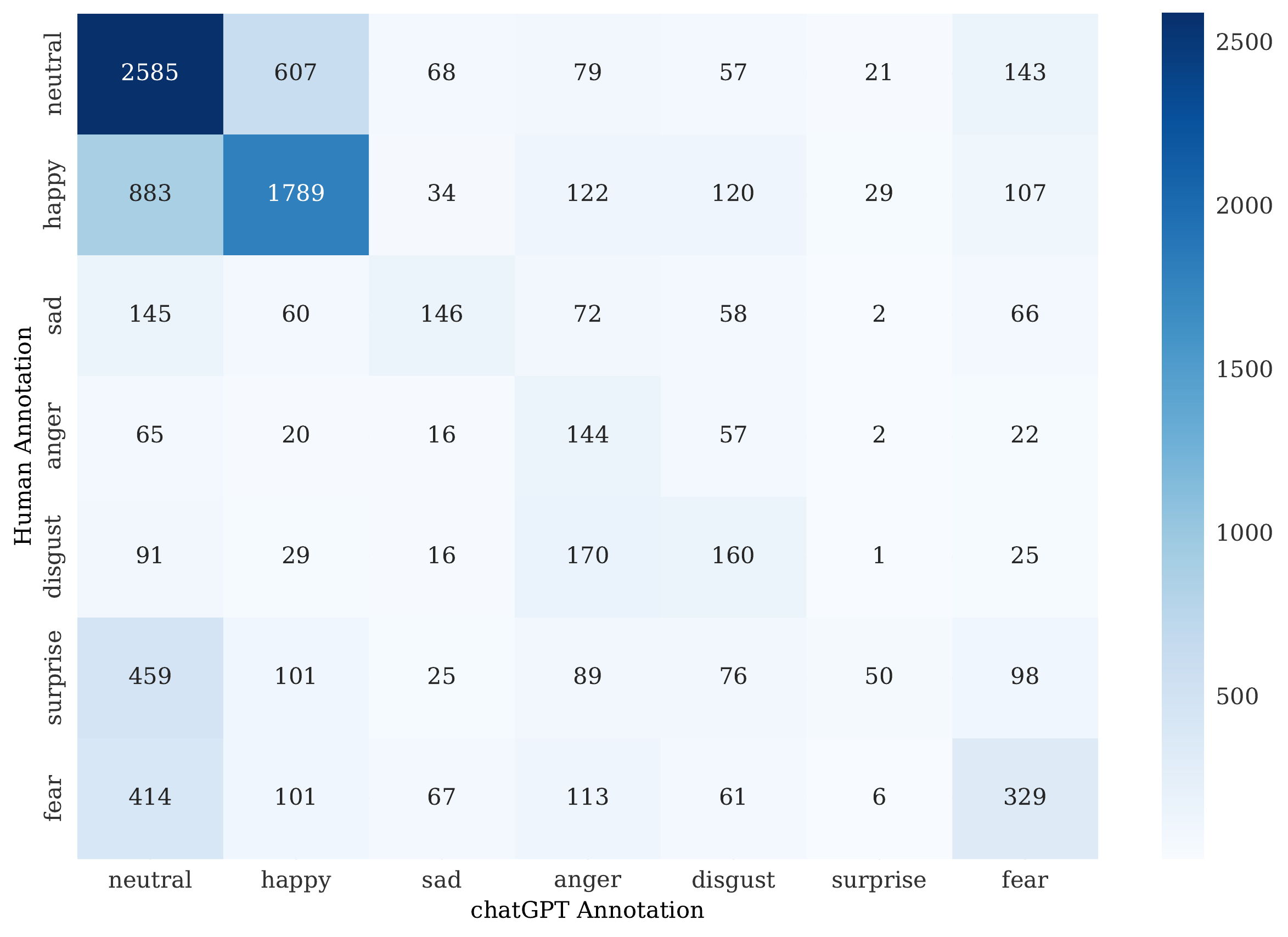}

\caption{Human vs. chatGPT Annotations}\label{fig:annotations}
\end{center}
  \begin{flushleft}
	\footnotesize{Notes: Confusion matrix of human vs. chatGPT annotations for our StockTwits training data.}
  \end{flushleft}
\end{figure}

We contrast human and ChatGPT annotations for our seven basic emotions. In Figure \ref{fig:annotations}, each row corresponds to the human annotations, and each column represents the annotations made by ChatGPT. Human and chatGPT annotations agree most often on neutral and happy emotions, with respective counts of 2585 and 1789 agreements. However, when it comes to more nuanced emotional categories such as sad, anger, disgust, surprise and fear, chatGPT and human annotations began to diverge. 

One trend evident in the data is ChatGPT's inclination towards categorizing instances into the 'neutral' category, which could stem from difficulties in parsing more nuanced emotional undertones. Alternatively, it might indicate a discrepancy in human labeling, wherein genuinely neutral texts might have been mistakenly assigned to other categories. In contrast, the model seems to under-predict 'surprise' and 'fear' emotions. The disparity observed in the classification of different emotional categories serves to highlight the inherent challenges of emotion classification tasks, especially when it comes to deciphering complex or closely-related emotions. This underlines the necessity of evaluating our models not only on the seven identified categories but also on a broader scale encompassing negative, neutral, and positive emotions.

\section{Natural Language Processing to Extract Emotions}\label{app:NLP}

\subsection{Text Processing and Model Fitting}
Before processing the text, we apply several preprocessing steps. First, we remove images, hyperlinks, and tags from the text. Then, we convert the text to lowercase, fix contractions (e.g., "i've" to "i have"), and correct common misspellings using SymSpell.\footnote{To correct misspellings, we create a vocabulary from our combined StockTwits and Twitter data, along with word frequencies. We remove stock tickers, emojis, emoticons, and words found in the NLTK corpus. Next, we check if the remaining words are in the SymSpell dictionary. If a word is found, no correction is necessary. For words not in the SymSpell dictionary, we segment them (e.g., "ilike" becomes "i like"). If the corrected segment differs from the original word only in spacing and both words are in the SymSpell dictionary, we use the segmentation.}

We replace numbers with "<number>", stock tickers with "<ticker>", company names with "<company>", user names with "<user>", and all other unknown tokens with "<unknown>". The messages are then tokenized, converting words to numbers, and split into sentences using HuggingFace.

For the emotion classification task, we employ DistilBERT, a powerful model capable of capturing word order, word usage, and local context. The model is trained to minimize prediction error using a categorical cross-entropy loss function. For a more detailed understanding of the Transformer architecture, please refer to Section 10.7 of \cite{zhang2021dive}.

\subsection{Model Implementation}

In our emotion classification task, we use a dataset consisting of approximately 250,000 messages, with each message containing 64 words. To train our model, we employ a fine-tuned DistilBERT model, employing a batch size of 128, a learning rate of 2e-5, and an early-stopping parameter of 1.

Due to the substantial number of parameters in deep learning models and the computational demands of backpropagation, traditional computing resources such as CPUs are not practical for our purposes. Therefore, we leverage the computational power of GPUs, specifically the NVIDIA GeForce RTX3080 GPU, to train our models efficiently. GPUs have demonstrated their effectiveness in deep learning tasks (see \cite{schmidhuberdeep}).

For our analysis, we rely on Python 3.8, utilizing essential libraries such as NumPy (\cite{walt2011numpy}), pandas (\cite{mckinney2010data}), and matplotlib (\cite{hunter2007matplotlib}). To fine-tune the DistilBERT model, we make use of the HuggingFace library (\cite{wolf2019huggingface}), which operates on top of Facebook Torch (\cite{NEURIPS2019_9015}).

\subsection{Examples of Messages \& Outputs}

We provide examples of our model's predictions in Table \ref{tab:twit_examples}. 

\input{Tables/twit_examples}

\subsection{Model Explanations}

To uncover the associations between words and the predictions of our DistilBERT model, we employ SHapley Additive exPlanations (SHAP), which is a comprehensive framework for interpreting predictions (\cite{SHAP}). SHAP utilizes game theory principles to assign local importance values to each feature (word) for a specific prediction. By combining these local explanations, we can generate global explanations by averaging the SHAP values word-wise. This enables us to compare words based on their average SHAP values, where higher values indicate higher predicted probabilities towards a class, and lower values indicate lower probabilities.

To conduct the SHAP analysis, we utilize a dataset of 10,000 StockTwits messages that have been manually tagged. Table \ref{tab:shap_stocktwits} presents the average SHAP values obtained from our StockTwits model. A brief examination of the table confirms that our model establishes meaningful relationships between words and emotions. For example, the word "surprise" (Panel B, Column 2) is associated with an average decrease of 20 percentage points in the predicted probability of the neutral class. 

\input{Tables/shap_stocktwits}

\clearpage 

\section{Variable Definitions and Sources}

Table \ref{tab:variable_definitions} defines the variables used in the analyses. 

\section{Sensitivity Analysis}\label{sec:sensitivity}
Next, we show that our results are robust, capture investor emotions intuitively, and their predictive power diminishes over time.

\subsection{Alternative Classification}

Perhaps the most important part of our robustness checks, we construct alternative emotion variables based on a model trained on a Twitter emotion metadata compiled by other researchers. We explore our Twitter based model in Column (1-3) of Table \ref{tab:pre_market_robustness}. Most our findings remain significant, which provides strong support for the relationship between investor emotions and daily price movements. 

\subsection{Alternative Weighting}

For our prior results, we weighted messages by log(1+followers). As a validity check, we consider an alternative weighting scheme. In particular, we investigate abandoning the weighting scheme, and hence, messages are weighted equally (Column (4-6) of Table \ref{tab:pre_market_robustness}). The results are largely unaltered under these alternative specifications.

\subsection{Contemporaneous Emotions \& Prices}

To provide further support that our emotion measures are capturing investor emotions accurately, we also estimate our main regression with contemporaneous (trading hour) emotion variables. We expect to find that when daily returns are high, people should be happy, and hence valence should be high as well. Table \ref{tab:market_ret} provides support for this. Interestingly, while pre-market sentiment-valence point estimate were comparable, the point estimates for trading hour valence are significantly higher than the StockTwits provided sentiment measure's, and valence explains an additional 5.39 percentage point (37.5\% increase) in the variation of daily returns beyond StockTwits sentiment. One potential explanation for this disparity is the reactive chat-like messages, emojis and emoticons ignored by the sentiment model that capture investor excitement in response to price movements. We abstain from analyzing the contemporaneous effects further, since these could be reactive and not predictive (i.e., we do not know whether emotion leads or lags price movements).

\subsection{Relationship with Self-Tagged Sentiment}
In Table \ref{tab:corr_matrix} we saw a positive correlation between our valence measure and the sentiment score provided by StockTwits. As mentioned earlier, we observe sentiment indicators as tagged by the user (bullish, bearish, or unclassified) for each message. In particular, we found that 9.6M messages are tagged out of our sample of 88M messages. We now dive deeper and explore the relationship between self-tagged sentiment and our emotion variables.  To do so, we restrict our sample to bullish and bearish messages and run regressions of sentiment on the emotion variables at the message level. Column (1) shows that valence and self-tagged sentiment are positively related, yet the large magnitude of the constant variable indicate that sentiment carries information beyond what our valence measure contains. This relationship remains significant even after controlling for firm and date fixed effects (Column (3)). We also looked at the relationship componentwise. For this we omitted the neutral group (to avoid perfect multi-collinearity), and hence this groups serves as the benchmark. Emotions behave mostly as expected (Column (2) and Column (4)), with one notable exception. Anger, unlike the other emotions with negative valence, has a small positive coefficient. We conjecture that this is driven by some obscene words, such as the ``f" word, which can express both anger and excitement.

\subsection{Predicting Several Days Ahead}

Table \ref{tab:pre_market_reversal} explores the impact on pre-market emotions today on returns the following four days. As expected, we find that the predictive power of investor emotions diminishes.

\input{Tables/variable_definitions}
\input{Tables/pre_market_ret_sensitivity}
\input{Tables/market_ret}

\input{Tables/sentiment_emotion}

\input{Tables/pre_market_reversal}

\clearpage 

%% file: Tables/twit_examples.tex
\begin{table}[htbp]\centering
\begin{threeparttable}
\footnotesize 
\caption{Examples of StockTwits Model Outputs}\label{tab:twit_examples}
\begin{tabular}{llllllll}
Text                                         & Neutral & Happy  & Sad    & Anger  & Disgust & Surprise & Fear   \\ \hline 
Financial markets have been uneventful.      & 79.2\%  & 1.9\%  & 5.3\%  & 1.7\%  & 7.4\%   & 1.3\%    & 3.1\%  \\
Today has been such a nice day :).           & 0.5\%   & 99.0\% & 0.1\%  & 0.0\%  & 0.0\%   & 0.2\%    & 0.1\%  \\
Been a long time since i felt so awful :(. & 0.4\%   & 4.1\%  & 81.0\% & 1.1\%  & 4.5\%   & 3.6\%    & 5.4\%  \\
You freaking idiots! Stop selling!!          & 0.6\%   & 1.2\%  & 3.5\%  & 63.8\% & 26.0\%  & 2.4\%    & 2.4\%  \\
These nasty politicians gotta go!            & 2.0\%   & 7.0\%  & 1.6\%  & 12.6\% & 63.8\%  & 0.9\%    & 12.0\% \\
WTf is going on rn??                         & 2.7\%   & 0.3\%  & 0.7\%  & 0.8\%  & 0.6\%   & 93.7\%   & 1.2\%  \\
Pretty choppy lately!                        & 22.5\%  & 11.9\% & 23.0\% & 3.9\%  & 6.4\%   & 1.9\%    & 30.4\% \\ \hline \hline 
\end{tabular}
\begin{tablenotes}
\footnotesize 
\item Notes: Examples of inputs (text) and outputs (i.e., neutral, happy, sad, anger, disgust, surprise, fear) for our StockTwits based emotion classificaiton model.
\end{tablenotes}
\end{threeparttable}
\end{table}

%% file: Tables/shap_stocktwits.tex
\begin{sidewaystable}[htbp]\centering
\caption{Shap Values: StockTwits Model}\label{tab:shap_stocktwits}
\begin{threeparttable}
\scriptsize
\begin{tabular}{llllllllllllll}
\multicolumn{2}{c}{Neutral}  & \multicolumn{2}{c}{Happy}  & \multicolumn{2}{c}{Sad}  & \multicolumn{2}{c}{Anger} & \multicolumn{2}{c}{Disgust} & \multicolumn{2}{c}{Surprise}  & \multicolumn{2}{c}{Fear}  \\ \hline 
& & & & & & & & & & & & & \\ [\dimexpr-\normalbaselineskip+2pt]
\multicolumn{14}{l}{\textbf{Panel A: Positive Predictors}} \\
& & & & & & & & & & & & & \\ [\dimexpr-\normalbaselineskip+2pt]
scala       & 0.343 & excellent  & 0.524 & sadly         & 0.448 & a**hole   & 0.222 & mole      & 0.371 & surprise   & 0.627 & scary    & 0.374 \\
rental      & 0.201 & amazing    & 0.481 & disappointed  & 0.388 & f***      & 0.220 & bastards  & 0.367 & surprised  & 0.565 & scared   & 0.360 \\
snaps       & 0.167 & happy      & 0.476 & sad           & 0.385 & su***     & 0.187 & stink     & 0.306 & wow        & 0.557 & fear     & 0.351 \\
transaction & 0.153 & beautiful  & 0.467 & sma           & 0.290 & iating    & 0.151 & a**hole   & 0.281 & curious    & 0.510 & scare    & 0.333 \\
incoming    & 0.147 & great      & 0.463 & sorry         & 0.247 & stupid    & 0.150 & ugly      & 0.242 & why        & 0.327 & nervous  & 0.318 \\
payroll     & 0.142 & awesome    & 0.443 & heartbreak    & 0.240 & bastards  & 0.149 & garbage   & 0.223 & wonder     & 0.302 & panic    & 0.316 \\
backward    & 0.142 & loving     & 0.434 & bleeding      & 0.226 & idiot     & 0.141 & idiots    & 0.190 & whoa       & 0.301 & worried  & 0.261 \\
cooke       & 0.135 & promising  & 0.423 & hurts         & 0.221 & f***ing   & 0.128 & savage    & 0.170 & strange    & 0.280 & concern  & 0.175 \\
upgraded    & 0.134 & impressive & 0.420 & unfortunately & 0.184 & sh*t      & 0.125 & hacker    & 0.164 & weird      & 0.172 & worry    & 0.160 \\
duction     & 0.122 & glad       & 0.414 & sober         & 0.184 & bi**h     & 0.122 & lth       & 0.162 & insane     & 0.159 & collapse & 0.130 \\
transcript  & 0.120 & excited    & 0.412 & crushed       & 0.183 & s***      & 0.119 & coward    & 0.158 & holy       & 0.150 & mort     & 0.123 \\
ved         & 0.119 & fun        & 0.407 & regret        & 0.175 & hate      & 0.117 & stupidity & 0.155 & tf         & 0.130 & scar     & 0.122 \\
files       & 0.118 & wonderful  & 0.399 & missed        & 0.174 & hacker    & 0.110 & agger     & 0.147 & crazy      & 0.126 & litter   & 0.121 \\
tanker      & 0.118 & bliss      & 0.389 & hurt          & 0.171 & stupidity & 0.104 & stupid    & 0.145 & incredible & 0.125 & bien     & 0.120 \\
amin        & 0.116 & ;)         & 0.352 & sinking       & 0.170 & d*mn      & 0.102 & idiot     & 0.143 & heck       & 0.123 & trouble  & 0.117 \\  
\multicolumn{14}{l}{\textbf{Panel B: Negative Predictors}} \\
& & & & & & & & & & & & & \\ [\dimexpr-\normalbaselineskip+2pt]
surprise  & -0.200 & fake         & -0.113 & curious    & -0.055 & scared     & -0.036 & wonderful       & -0.041 & boring    & -0.052 & beautiful & -0.073 \\
ugly      & -0.200 & wow          & -0.114 & happy      & -0.057 & awesome    & -0.036 & awesome         & -0.041 & nice      & -0.053 & emoji{person-surfing}      & -0.073 \\
beautiful & -0.204 & strange      & -0.115 & beautiful  & -0.058 & beautiful  & -0.037 & bliss           & -0.041 & loving    & -0.054 & great     & -0.073 \\
curious   & -0.207 & mole         & -0.117 & incredible & -0.058 & curious    & -0.039 & excited         & -0.043 & awesome   & -0.056 & happy     & -0.080 \\
great     & -0.208 & a**hole      & -0.117 & excited    & -0.059 & impressive & -0.040 & congratulations & -0.045 & beautiful & -0.056 & fun       & -0.081 \\
su**      & -0.214 & sad          & -0.118 & promising  & -0.059 & wonderful  & -0.040 & scare           & -0.046 & great     & -0.057 & wow       & -0.083 \\
awful     & -0.216 & surprise     & -0.118 & awesome    & -0.059 & scala      & -0.041 & nervous         & -0.047 & glad      & -0.059 & amazing   & -0.084 \\
mole      & -0.217 & garbage      & -0.118 & amazing    & -0.066 & excited    & -0.041 & scared          & -0.049 & sadly     & -0.060 & nice      & -0.085 \\
hacker    & -0.219 & bleeding     & -0.122 & wow        & -0.076 & loving     & -0.042 & amazing         & -0.049 & promising & -0.061 & loving    & -0.087 \\
su***     & -0.223 & stupidity    & -0.127 & surprised  & -0.080 & nding      & -0.043 & worried         & -0.049 & hilarious & -0.061 & excellent & -0.087 \\
amazing   & -0.226 & stupid       & -0.131 & scala      & -0.080 & amazing    & -0.043 & scala           & -0.050 & trouble   & -0.068 & uber      & -0.088 \\
nice      & -0.241 & stink        & -0.132 & bliss      & -0.081 & surprised  & -0.047 & surprised       & -0.055 & happy     & -0.074 & awesome   & -0.098 \\
excellent & -0.241 & sadly        & -0.147 & wonderful  & -0.082 & surprise   & -0.048 & surprise        & -0.056 & excellent & -0.075 & bliss     & -0.099 \\
bastards  & -0.249 & disappointed & -0.151 & surprise   & -0.090 & wow        & -0.050 & panic           & -0.056 & wonderful & -0.075 & surprised & -0.109 \\
a**hole   & -0.249 & ugly         & -0.158 & savage     & -0.158 & acle       & -0.074 & wow             & -0.060 & scala     & -0.081 & surprise  & -0.115 \\
 \hline \hline 
\end{tabular}
\begin{tablenotes}
\scriptsize 
\item Notes: Average SHAP values evaluated on 10,000 hand-tagged StockTwits messages for the StockTwits based model. Words reported, followed by their corresponding average SHAP values, are the 15 most important words for each class. For each class separately, Panel A reports the top 15 words that are positively, while Panel B shows the top 15 words that are negatively associated with predictions. We restrict our SHAP analysis for words that appear at least twice in the SHAP sample. For instance, the word surprise (Panel B Column 2) is associated with a 20 percentage point decrease in the neutral predicted probability on average.
\end{tablenotes}
\end{threeparttable}
\end{sidewaystable}

%% file: Tables/variable_definitions.tex
\begin{center}
\onehalfspacing
\footnotesize 
\begin{ThreePartTable}

\begin{longtable}[H]{p{1.5in}p{4in}p{0.75in}}
\caption{Variable Definitions}\label{tab:variable_definitions} \\

Variable & Definition & Source \\ \hline 
\endfirsthead

\multicolumn{3}{c}%
{{\bfseries \tablename\ \thetable{} -- continued from previous page}} \\
Variable & Definition & Source \\ \hline 
& & \\[\dimexpr-\normalbaselineskip+5pt]
\endhead

\endlastfoot
Open-Close Return & Difference between daily closing and open price, normalized by the open price. & CRSP \\
Close-Open Return & Difference between previous day's closing price and current open price, normalized by the previous day's closing price. & CRSP \\
Market Cap$_{-1}$  & Natural logarithm of market value of equity ($\log$(1+CSHOC$*$PRCCD)). & CRSP \\
Volatility & Standard deviation of daily returns during from 183 days prior up to a day before. & CRSP \\
Short Interest & Shares short divided by shares outstanding. Bi-weekly frequency. & Compustat \\
Emotion & Each message is classified by a many-to-one deep learning model into one of the seven categories (i.e., neutral, happy, sad, anger, disgust, surprise, fear), so that the corresponding probabilities sum up to 1. For each emotions separately, we then take the weighted average of these probabilities from 4pm the previous day until 9.29am the trading day and the weights correspond to the number of followers of the user 1+$\log$(1 + \# of Followers).  & StockTwits \\
& & \\ 
 \hline \hline 
\end{longtable}
\end{ThreePartTable}
\end{center}

%% file: Tables/pre_market_ret_sensitivity.tex
\begin{sidewaystable}[htbp]\centering
\footnotesize
\begin{threeparttable}
\def\sym#1{\ifmmode^{#1}\else\(^{#1}\)\fi}
\caption{Robustness: Pre-Market Emotions and Price Movements \label{tab:pre_market_robustness}}
\begin{tabular}{l*{6}{c}}
\hline\hline
                    &\multicolumn{1}{c}{(1)}&\multicolumn{1}{c}{(2)}&\multicolumn{1}{c}{(3)}&\multicolumn{1}{c}{(4)}&\multicolumn{1}{c}{(5)}&\multicolumn{1}{c}{(6)}\\
                   &\multicolumn{3}{c}{Twitter Model}&\multicolumn{1}{c}{Alternative Weighting}\\
\hline
 &                     &                     &                     &   & &             \\[\dimexpr-\normalbaselineskip+2pt]
 StockTwits Sentiment     &      0.0055\sym{***}&      0.0015\sym{***}&      0.0309\sym{***}&      0.0053\sym{***}&      0.0019\sym{***}&      0.0215\sym{***}\\
                    &    (0.0004)         &    (0.0004)         &    (0.0046)         &    (0.0003)         &    (0.0004)         &    (0.0048)         \\
Valence             &      0.0048\sym{***}&      0.0033\sym{***}&      0.0001         &      0.0060\sym{***}&      0.0020\sym{***}&      0.0158\sym{***}\\
                    &    (0.0006)         &    (0.0006)         &    (0.0053)         &    (0.0006)         &    (0.0007)         &    (0.0044)         \\
 &                     &                     &                     &   & &             \\[\dimexpr-\normalbaselineskip+2pt]

\hline
 &                     &                     &                     &   & &             \\[\dimexpr-\normalbaselineskip+2pt]
S\&P Super Composite Firms      &                     &           X         &                     &                     &           X         &                     \\
At least 100 messages&                     &                     &           X         &                     &                     &           X         \\ \hline 
 &                     &                     &                     &   & &             \\[\dimexpr-\normalbaselineskip+2pt]
$\sigma_{y, within}$&      0.0480         &      0.0309         &      0.0673         &      0.0480         &      0.0309         &      0.0673         \\
Observations        & 454138       & 163006       &  53803       & 454138       & 163006       &  53803       \\
$R^2$               &      0.1055         &      0.1444         &      0.1741         &      0.1056         &      0.1443         &      0.1743         \\
\hline\hline
\end{tabular}
\begin{tablenotes}
\footnotesize 
\item Notes: This table considers the robustness of the relationship between pre-market emotions and daily price movements. Columns (1-3) uses emotion variables constructed based on a model trained on the emotion meta-data compiled by other researchers. Columns (4-6) abandons the weighting scheme used in our analysis, and weighs each posts equally. The dependent variable is daily returns, computed as the difference between the closing and the opening price, divided by the opening price. All specifications include firm and date fixed effects, close-open returns, lag open-close returns, past 20 trading days return and volatility (as defined in Variable Definitions). Robust standard errors clustered at the industry and date levels are in parentheses. We use the Fama-French 12-industry classification. \sym{*} \(p<0.10\), \sym{**} \(p<0.05\), \sym{***} \(p<0.01\). Continuous variables winsorized at the 0.1\% and 99.9\% level to mitigate the impact of outliers. We report the within-firm, detrended (demeaned) standard deviation of the dependent variable. 
\end{tablenotes}
\end{threeparttable}
\end{sidewaystable}

%% file: Tables/market_ret.tex
\begin{table}[htbp]\centering
\footnotesize
\begin{threeparttable}
\def\sym#1{\ifmmode^{#1}\else\(^{#1}\)\fi}
\caption{Market Emotions and Price Movements \label{tab:market_ret}}
\begin{tabular}[lcccc]{p{1.5in}p{1in}p{1in}p{1in}p{1in}}
\hline\hline
 &                     &                     &                     &                \\[\dimexpr-\normalbaselineskip+2pt]
                    &\multicolumn{1}{c}{(1)}&\multicolumn{1}{c}{(2)}&\multicolumn{1}{c}{(3)}&\multicolumn{1}{c}{(4)}\\
\hline
 &                     &                     &                     &                                     \\[\dimexpr-\normalbaselineskip+2pt]
StockTwits Sentiment     &      0.0618\sym{***}&      0.0256\sym{***}&      0.0141\sym{***}&      0.0429\sym{***}\\
                    &    (0.0005)         &    (0.0005)         &    (0.0005)         &    (0.0041)         \\
 &                     &                     &                     &                                     \\[\dimexpr-\normalbaselineskip+2pt]
Valence             &                     &      0.0926\sym{***}&      0.0545\sym{***}&      0.3468\sym{***}\\
                    &                     &    (0.0006)         &    (0.0007)         &    (0.0037)         \\
\hline
                     &                     &                     &                    &                       \\[\dimexpr-\normalbaselineskip+2pt]
S\&P Super Composite Firms&                     &                     &           X         &                     \\
At least 100 messages&                     &                     &                     &           X         \\ \hline 
                     &                     &                     &                    &                       \\[\dimexpr-\normalbaselineskip+2pt]
$\sigma_{y, within}$&   0.0539         &      0.0539         &      0.0350         &      0.0773         \\
Observations        & 476376       & 476376       & 160335       &  78268       \\
$R^2$               &      0.1435         &      0.1974         &      0.2285         &      0.3324         \\
\hline\hline
\end{tabular}
\begin{tablenotes}
\footnotesize 
\item Notes: This table considers the relationship between emotions posted during market hours and daily price movements. The dependent variable is daily returns, computed as the difference between the closing and the opening price, divided by the opening price. All specifications include firm and date fixed effects, close-open returns, lag open-close returns, past 20 trading days return and volatility (as defined in Variable Definitions). Robust standard errors clustered at the industry and date levels are in parentheses. We use the Fama-French 12-industry classification. \sym{*} \(p<0.10\), \sym{**} \(p<0.05\), \sym{***} \(p<0.01\). Continuous variables winsorized at the 0.1\% and 99.9\% level to mitigate the impact of outliers. We report the within-firm, detrended (demeaned) standard deviation of the dependent variable. 
\end{tablenotes}
\end{threeparttable}
\end{table}

%% file: Tables/sentiment_emotion.tex
\begin{table}[htbp]\centering
\footnotesize
\begin{threeparttable}
\def\sym#1{\ifmmode^{#1}\else\(^{#1}\)\fi}
\caption{Relationship between Self-Tagged Sentiment and EmTract Emotions \label{tab:sentiment_emotion}}
\begin{tabular}[lcccc]{p{1.5in}p{1in}p{1in}p{1in}p{1in}}
\hline\hline
 &                     &                     &                     &                \\[\dimexpr-\normalbaselineskip+2pt]
                    &\multicolumn{1}{c}{(1)}&\multicolumn{1}{c}{(2)}&\multicolumn{1}{c}{(3)}&\multicolumn{1}{c}{(4)}\\
\hline
 &                     &                     &                     &                                     \\[\dimexpr-\normalbaselineskip+2pt]
Valence             &      0.1591\sym{***}&                     &      0.1394\sym{***}&                     \\
                    &    (0.0023)         &                     &    (0.0020)         &                     \\
 &                     &                     &                     &                                     \\[\dimexpr-\normalbaselineskip+2pt]
Happy               &                     &      0.0964\sym{***}&                     &      0.0692\sym{***}\\
                    &                     &    (0.0019)         &                     &    (0.0016)         \\
 &                     &                     &                     &                                     \\[\dimexpr-\normalbaselineskip+2pt]
Sad                 &                     &     -0.3677\sym{***}&                     &     -0.3488\sym{***}\\
                    &                     &    (0.0054)         &                     &    (0.0050)         \\
 &                     &                     &                     &                                     \\[\dimexpr-\normalbaselineskip+2pt]
Anger               &                     &      0.0657\sym{***}&                     &      0.0271\sym{***}\\
                    &                     &    (0.0035)         &                     &    (0.0026)         \\
 &                     &                     &                     &                                     \\[\dimexpr-\normalbaselineskip+2pt]
Disgust             &                     &     -0.3438\sym{***}&                     &     -0.3322\sym{***}\\
                    &                     &    (0.0035)         &                     &    (0.0027)         \\
 &                     &                     &                     &                                     \\[\dimexpr-\normalbaselineskip+2pt]
Surprise            &                     &     -0.0208\sym{***}&                     &     -0.0370\sym{***}\\
                    &                     &    (0.0015)         &                     &    (0.0013)         \\
 &                     &                     &                     &                                     \\[\dimexpr-\normalbaselineskip+2pt]
Fear                &                     &     -0.2266\sym{***}&                     &     -0.2020\sym{***}\\
                    &                     &    (0.0040)         &                     &    (0.0035)         \\
 &                     &                     &                     &                                     \\[\dimexpr-\normalbaselineskip+2pt]
Constant            &      0.8226\sym{***}&      0.8676\sym{***}&      0.8251\sym{***}&      0.8761\sym{***}\\
                    &    (0.0024)         &    (0.0023)         &    (0.0003)         &    (0.0007)         \\
\hline
                     &                     &                     &                    &                       \\[\dimexpr-\normalbaselineskip+2pt]
Date Fixed Effects  &                     &                     &           X         &           X         \\
Firm Fixed Effects  &                     &                     &           X         &           X         \\ \hline 
                     &                     &                     &                    &                       \\[\dimexpr-\normalbaselineskip+2pt]
$\sigma_{y, within}$&      0.3641         &      0.3641         &      0.3436         &      0.3436         \\
Observations        &9636817      &9636817      &9636673      &9636673      \\
$R^2$               &      0.0566         &      0.0641         &      0.1523         &      0.1598         \\
\hline\hline
\end{tabular}
\begin{tablenotes}
\footnotesize 
\item Notes: This table considers the relationship between emotions and self-tagged sentiment. The dependent variable is self-tagged sentiment, provided by the author of the post. Column (3) and (4) include firm and date fixed effects. Robust standard errors clustered at the calendar date level are in parentheses. \sym{*} \(p<0.10\), \sym{**} \(p<0.05\), \sym{***} \(p<0.01\). 
\end{tablenotes}
\end{threeparttable}
\end{table}

%% file: Tables/pre_market_reversal.tex
\begin{table}[htbp]\centering
\footnotesize
\begin{threeparttable}
\def\sym#1{\ifmmode^{#1}\else\(^{#1}\)\fi}
\caption{Pre-Market Emotions and Subsequent Price Movements \label{tab:pre_market_reversal}}
\begin{tabular}[lcccc]{p{1.5in}p{1in}p{1in}p{1in}p{1in}}
\hline\hline
                    &\multicolumn{1}{c}{(1)}&\multicolumn{1}{c}{(2)}&\multicolumn{1}{c}{(3)}&\multicolumn{1}{c}{(4)}\\
                    &\multicolumn{1}{c}{Ret$_{t+1}$}&\multicolumn{1}{c}{Ret$_{t+2}$}&\multicolumn{1}{c}{Ret$_{t+3}$}&\multicolumn{1}{c}{Ret$_{t+4}$}\\
\hline
 &                     &                     &                     &                \\[\dimexpr-\normalbaselineskip+2pt]
StockTwits Sentiment     &      0.0010\sym{***}&      0.0005         &      0.0004         &      0.0005\sym{*}  \\
                    &    (0.0003)         &    (0.0003)         &    (0.0003)         &    (0.0003)         \\
    &                     &                     &                     &                \\[\dimexpr-\normalbaselineskip+2pt]                 
EmTract Valence             &     -0.0006         &     -0.0003         &     -0.0005         &     -0.0004         \\
                    &    (0.0005)         &    (0.0005)         &    (0.0005)         &    (0.0005)         \\

\hline
 &                     &                     &                     &                \\[\dimexpr-\normalbaselineskip+2pt]
$\sigma_{y, within}$&         0.0431         &      0.0418         &      0.0410         &      0.0405         \\
Observations        & 453979        & 453877        & 453796        & 453689        \\
$R^2$               &      0.1108         &      0.1141         &      0.1180         &      0.1159         \\
\hline\hline
\end{tabular}
\begin{tablenotes}
\footnotesize 
\item Notes: This table investigates the predictive power of investor emotions on price movements one to four days into the future. The dependent variable is daily returns, computed as the difference between the closing and the opening price, divided by the opening price at $t+1$ through $t+4$. All specifications include firm and date fixed effects, close-open returns, open-close returns at $t-1$, past 20 trading days return and volatility (as defined in Variable Definitions). Robust standard errors clustered at the industry and date levels are in parentheses. We use the Fama-French 12-industry classification. \sym{*} \(p<0.10\), \sym{**} \(p<0.05\), \sym{***} \(p<0.01\). Continuous variables winsorized at the 0.1\% and 99.9\% level to mitigate the impact of outliers. We report the within-firm, detrended (demeaned) standard deviation of the dependent variable. 
\end{tablenotes}
\end{threeparttable}
\end{table}

%% file: Sections/OnlineAppendix.tex
\onehalfspacing

\appendix
\clearpage

\counterwithin{figure}{section}
\counterwithin{table}{section}
\addcontentsline{toc}{section}{Appendices}

\section*{Online Appendix*: EmTract: Extracting Emotions from Social Media}\label{app:online_app}

\section{Alternative Open-Source Packages}\label{app:user_guide}

\cite{mishra2021deep} use the Valence Aware Dictionary and Sentiment Reasoner (VADER) (see \cite{elbagir2019twitter}) for sentiment labeling as inputs for a deep learning sentiment model. We prefer our NLP models over dictionary-based methods for emotion classification due to their probabilistic class assessments and trained word-embeddings that learn co-occurring words, thus providing valuable inference information even for words not in our training sample.

We compared our models with an open-source dictionary-based emotion library, Text2Emotion, which classifies text into five classes: happy, sad, fear, anger, surprise. Text2Emotion showed significantly poorer and slower performance on our hand-tagged sample, accurately classifying only 25.2\% of our posts, with a much higher loss (20.46). Also, Text2Emotion took 12 minutes to compute emotions for our sample, while our model did so in seconds. We have made a user guide for our Python package, EmTract, available at \href{https://github.com/dvamossy/EmTract}{https://github.com/dvamossy/EmTract}..

\section{Messages as Indicators for Emotion}

Our emotion measure should reflect the genuine sentiment of investors, ruling out potential stock market manipulation via false opinions. Despite potential manipulation concerns, our data seems robust due to anecdotal evidence of users posting to attract followers or find employment, thus incentivized to share honest stock opinions. Furthermore, we focus on S\&P 500 firms, whose large market caps make it unlikely for individual investors to influence prices.

\section{Measuring Information Content}

We also differentiate the emotional state of messages based on their relevance to finance or general chat. We use a dictionary to identify posts containing information related to earnings, firm fundamentals, or stock trading, classifying them as "finance." Our dictionary is \href{https://github.com/dvamossy/EmTract/tree/main/emtract/data}{available on our GitHub repository}. Messages providing original information are distinguished from those disseminating existing information, with a message considered original if it isn't a retweet and doesn't include a hyperlink.  
\section{StockTwits Activity, Correlations, and Sample Distributions}

\begin{figure}[htbp] 

\begin{center}
	\subfloat[]{\includegraphics[scale=0.3]{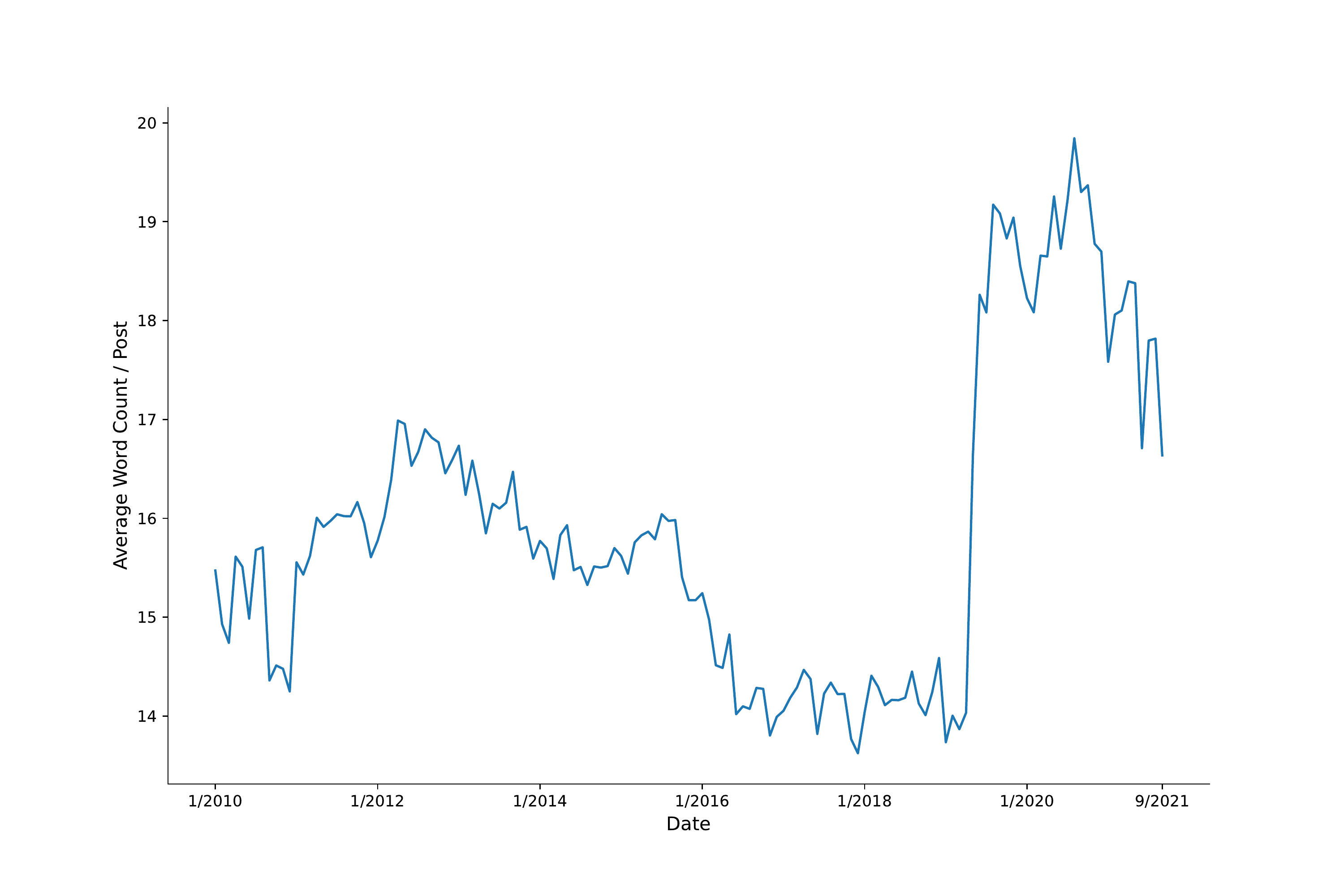}}
	\subfloat[]{\includegraphics[scale=0.325]{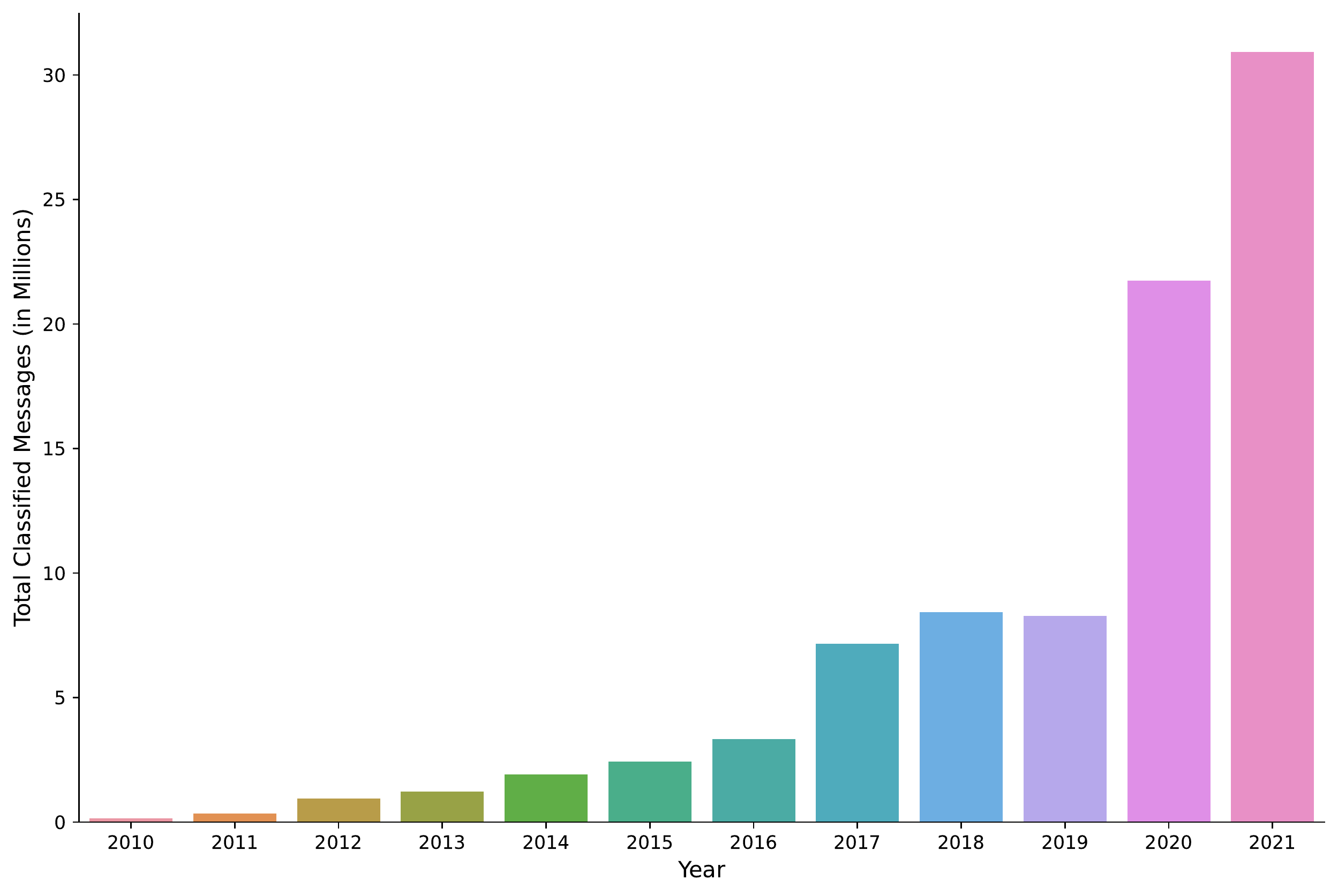}}

	\subfloat[]{\includegraphics[scale=0.325]{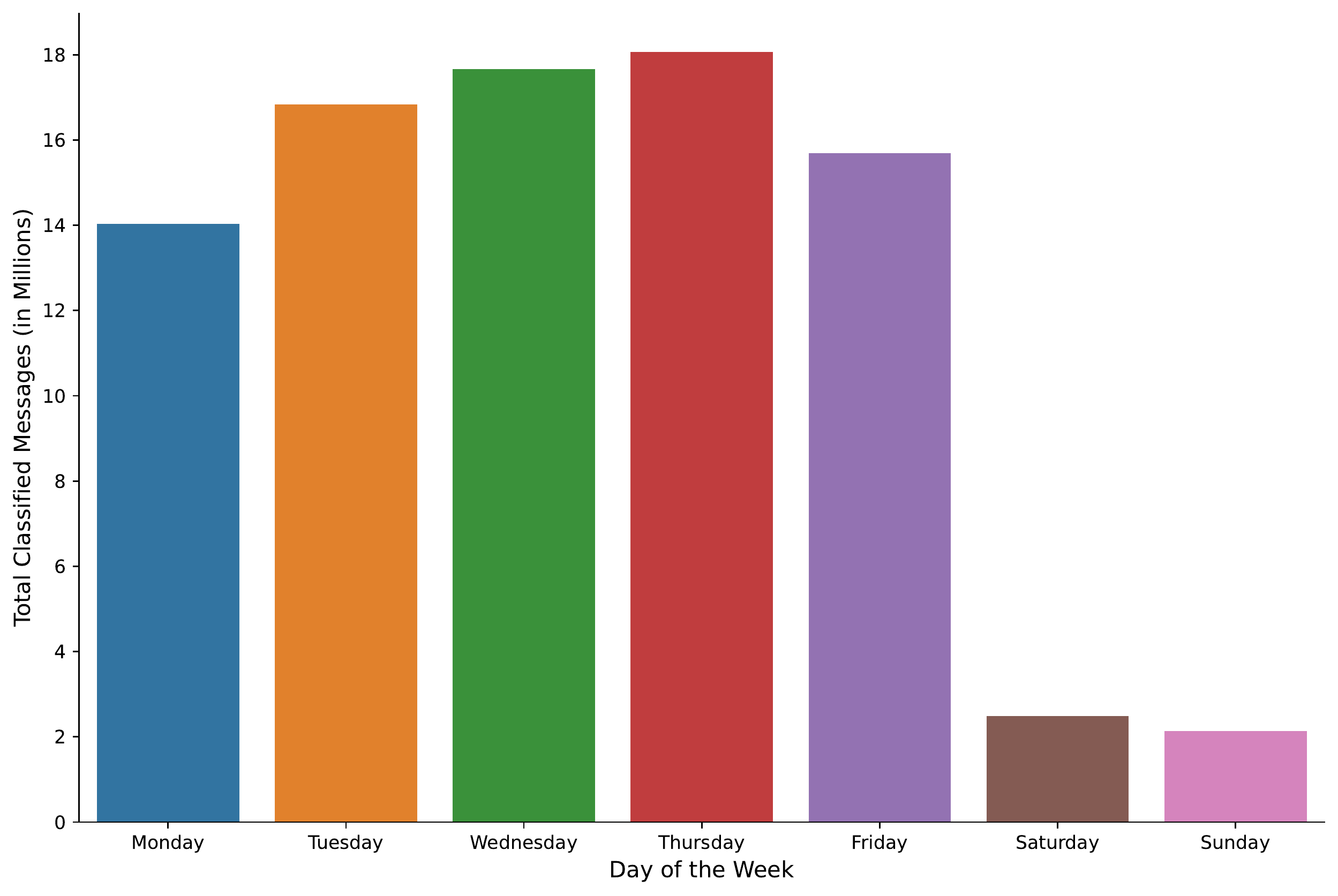}}
	\subfloat[]{\includegraphics[scale=0.325]{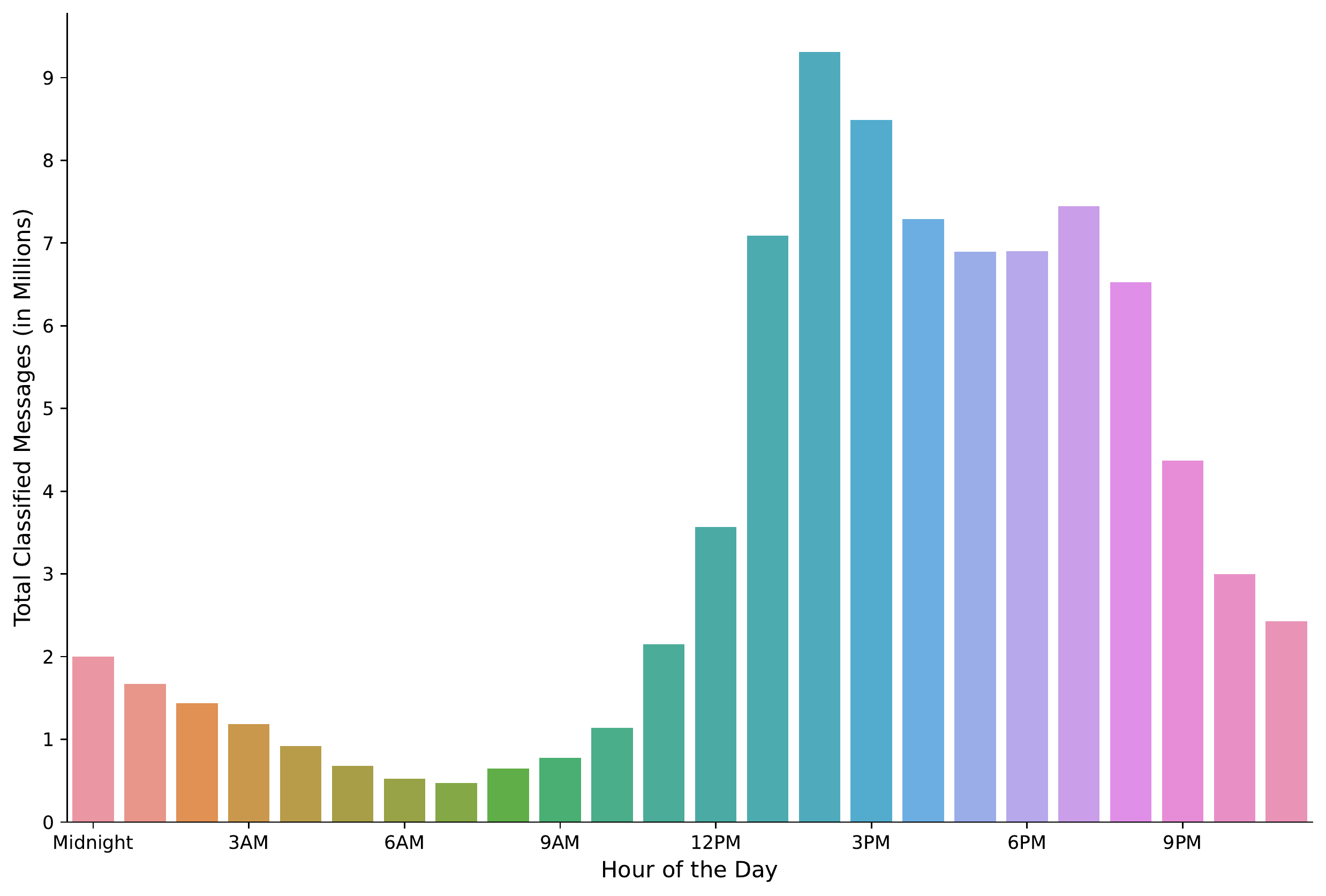}}

	\caption{Post Characteristics}\label{fig:post_distributions}
\end{center}
 
\begin{flushleft}
	\footnotesize{Notes: Includes our final pre-market and market sample. Panel (a) plots the average word count per post over time, while Panel (b) shows the number of posts over time in our data. Panel (c) portrays the day-of-the-week, while Panel (d) depicts the hour-of-the-day distribution of posts.}
 
\end{flushleft}

\end{figure}

Figure \ref{fig:post_distributions} shows message word count and frequency over time, revealing a spike in word count after the 140-to-280 character limit increase. With 97.5\% messages containing 64 or fewer words, this limit change likely doesn't affect emotion estimation. Message frequency experienced significant growth until 2017, plateaued, and surged in 2020. We account for these changes using time fixed effects. We also investigate the timing of investor posts. Figures \ref{fig:post_distributions} (c) and (d) show most posts are made during market open hours, suggesting real-time belief updating by investors.

Table \ref{tab:corr_matrix} presents pairwise correlations among analysis variables including our valence measure, StockTwits sentiment, daily return, and control variables. Our valence measure correlates significantly with daily returns and StockTwits sentiment, hinting at the predictive ability of aggregate emotions from posts on price movements. Table \ref{tab:twits_by_year} shows a sharp increase in StockTwits activity from 2010 to 2020, reflecting social media's growing popularity. Our coverage also expands in firm-day observations. Table \ref{tab:fama_fench} reveals our sample spans all 12 Fama-French industries, with a heavier presence in tech and healthcare firms, and fewer financial firms.

\section{Heterogeneous Effects: User Characteristic}

\cite{hong2004groups} show that a diverse group of intelligent decision makers reach reliably better decisions than a less diverse group of individuals with superior skills. We investigate this by segmenting our messages coming from traders with similar investment horizons (i.e., long-term, short-term), trading approaches (i.e., value, technical), and trading experiences (i.e., amateur \& intermediate, professional). 

We report heterogeneity across user types in Table \ref{tab:pre_market_types} and document a few interesting observations. In line with the value of diversity hypothesis, we find that the emotions of homogeneous groups are less informative in predicting daily price movements. In addition, the relationship between investor emotions and daily price movements is stable, and statistically significant in most specifications. We also find that emotions extracted from messages coming from professional users have smaller point estimates than messages authored by non-professional users. This is in stark contrast with the StockTwits provided sentiment measure, for which posts authored by professional and non-professional traders have almost identical point estimates. Last, we document that posts coming from short-term traders have slightly larger point estimates than posts authored by long-term traders. 

\input{Tables/correlations}
\input{Tables/summary_year_tab} 
\input{Tables/summary_industry_tab}
\input{Tables/pre_market_heterogeneity_ret}

\clearpage 

%% file: Tables/correlations.tex
\begin{sidewaystable}[htbp]\centering
\begin{threeparttable}
\footnotesize
\caption{Correlation Matrix}\label{tab:corr_matrix}
\def\sym#1{\ifmmode^{#1}\else\(^{#1}\)\fi}
\begin{tabular}{l*{7}{c}}
\hline\hline
                &\multicolumn{7}{c}{(1)}                                                                                                             \\
                &\multicolumn{7}{c}{}                                                                                                                \\
                & Valence       & StockTwits Sentiment        & Return$_{oc}$         &   Return$_{oc, -1}$          & Return$_{co}$         & Return$_{-20,-1}$        &Volatility$_{-183,-1}$         \\
\hline
Valence&            &                  &                  &                  &                  &                  &                  \\
                &                  &                  &                  &                  &                  &                  &                  \\
[1em]
StockTwits Sentiment &     0.32\sym{***}&            &                  &                  &                  &                  &                  \\
                &                  &                  &                  &                  &                  &                  &                  \\
[1em]
Return$_{oc}$   &    -0.01\sym{***}&     0.00         &            &                  &                  &                  &                  \\
                &                  &                  &                  &                  &                  &                  &                  \\
[1em]
Return$_{oc, -1}$        &     0.21\sym{***}&     0.10\sym{***}&    -0.03\sym{***}&            &                  &                  &                  \\
                &                  &                  &                  &                  &                  &                  &                  \\
[1em]
Return$_{co}$  &     0.17\sym{***}&     0.13\sym{***}&    -0.07\sym{***}&     0.03\sym{***}&            &                  &                  \\
                &                  &                  &                  &                  &                  &                  &                  \\
[1em]
Return$_{-20,-1}$  &     0.15\sym{***}&     0.03\sym{***}&    -0.02\sym{***}&     0.20\sym{***}&    -0.00         &            &                  \\
                &                  &                  &                  &                  &                  &                  &                  \\
[1em]
Volatility$_{-183,-1}$  &     0.23\sym{***}&     0.08\sym{***}&    -0.04\sym{***}&     0.02\sym{***}&     0.06\sym{***}&     0.16\sym{***}&            \\
                &                  &                  &                  &                  &                  &                  &                  \\
\hline
Observations    &   479463         &                  &                  &                  &                  &                  &                  \\
\hline\hline
\end{tabular}
\begin{tablenotes}
\scriptsize 
\item Notes: Pairwise correlation with Bonferroni corrections. Continuous variables winsorized at the 0.1\% and 99.9\% levels. \textit{t} statistics in parentheses. \sym{*} \(p<0.05\), \sym{**} \(p<0.01\), \sym{***} \(p<0.001\). Return$_{co}$ refers to return between the closing price at $t-1$ and the opening price at $t$. Return$_{oc}$ refers to the open-close return.
\end{tablenotes}
\end{threeparttable}
\end{sidewaystable}

%% file: Tables/summary_year_tab.tex
\begin{table}[htbp]\centering 
\footnotesize
\caption{Distribution of Non-Trading Hour Posts by Calendar Year \label{tab:twits_by_year}}
\def\sym#1{\ifmmode^{#1}\else\(^{#1}\)\fi}
\begin{tabular}{l*{2}{c}}
\hline\hline
& & \\[\dimexpr-\normalbaselineskip+2pt]
                    &\multicolumn{1}{c}{(1)}&\multicolumn{1}{c}{(2)}\\
                    &\multicolumn{1}{c}{Firm-Day Observations} &\multicolumn{1}{c}{Posts} \\
\hline
& & \\[\dimexpr-\normalbaselineskip+2pt]
\hline
2010                &        1337         &       39768         \\
                    &                     &                     \\
 & & \\[\dimexpr-\normalbaselineskip+2pt]
2011                &        3543         &      115710         \\
                    &                     &                     \\
 & & \\[\dimexpr-\normalbaselineskip+2pt]
2012                &        4857         &      295552         \\
                    &                     &                     \\
 & & \\[\dimexpr-\normalbaselineskip+2pt]
2013                &        6718         &      399211         \\
                    &                     &                     \\
 & & \\[\dimexpr-\normalbaselineskip+2pt]
2014                &       11285         &      692239         \\
                    &                     &                     \\
 & & \\[\dimexpr-\normalbaselineskip+2pt]
2015                &       17256         &      893803         \\
                    &                     &                     \\
 & & \\[\dimexpr-\normalbaselineskip+2pt]
2016                &       22734         &     1270722         \\
                    &                     &                     \\
 & & \\[\dimexpr-\normalbaselineskip+2pt]
2017                &       41840         &     2796966         \\
                    &                     &                     \\
 & & \\[\dimexpr-\normalbaselineskip+2pt]
2018                &       53472         &     3453727         \\
                    &                     &                     \\
 & & \\[\dimexpr-\normalbaselineskip+2pt]
2019                &       58645         &     3351474         \\
                    &                     &                     \\
 & & \\[\dimexpr-\normalbaselineskip+2pt]
2020                &      116338         &     9772462         \\
                    &                     &                     \\
 & & \\[\dimexpr-\normalbaselineskip+2pt]
2021                &      141438         &    14575580         \\
                    &                     &                     \\
 & & \\[\dimexpr-\normalbaselineskip+2pt]
Total               &      479463         &    37657214         \\
                    &                     &                     \\
\hline\hline
\end{tabular}
\end{table}

%% file: Tables/summary_industry_tab.tex
\begin{sidewaystable}[htbp]\centering
\begin{threeparttable}
\caption{Distribution of Non-Trading Hour Posts Based on Fama-French 12-Industry Classification\label{tab:fama_fench}}
\footnotesize 
\def\sym#1{\ifmmode^{#1}\else\(^{#1}\)\fi}
\begin{tabular}{l*{3}{c}}
\hline\hline
& & & \\[\dimexpr-\normalbaselineskip+2pt]
                    &\multicolumn{1}{c}{(1)}&\multicolumn{1}{c}{(2)}&\multicolumn{1}{c}{(3)}\\
Fama-French industry code (12 industries)                    &         CRSP (\%)&         Posts (\%)& Firm-Day (\%) \\
\hline
& & & \\[\dimexpr-\normalbaselineskip+2pt]
Consumer NonDurables -- Food, Tobacco, Textiles, Apparel, Leather, Toys&        4.01&        2.73&        2.99\\
Consumer Durables -- Cars, TV's, Furniture, Household Appliances&        2.30&        7.60&        3.35\\
Manufacturing -- Machinery, Trucks, Planes, Off Furn, Paper, Com Printing&        8.58&        3.01&        5.04\\
Oil, Gas, and Coal Extraction and Products&        3.13&        1.56&        2.91\\
Chemicals and Allied Products&        2.38&        1.34&        1.61\\
Business Equipment -- Computers, Software, and Electronic Equipment&       15.01&       22.62&       21.82\\
Telephone and Television Transmission&        1.75&        1.11&        1.75\\
Utilities           &        2.50&        0.31&        0.86\\
Wholesale, Retail, and Some Services (Laundries, Repair Shops)&        7.48&        5.92&        7.58\\
Healthcare, Medical Equipment, and Drugs&       15.13&       31.96&       33.21\\
Finance             &       24.01&        4.21&        8.23\\
Other -- Mines, Constr, BldMt, Trans, Hotels, Bus Serv, Entertainment&       13.72&       17.63&       10.66\\
\hline
Observations        &    10357038&    36518413&      471307\\
\hline\hline
\end{tabular}
\begin{tablenotes}
\scriptsize 
\item Notes: CRSP sample follows the same basic sample restrictions; active, traded on U.S. exchanges.
\end{tablenotes}
\end{threeparttable}
\end{sidewaystable}

%% file: Tables/pre_market_heterogeneity_ret.tex
\begin{sidewaystable}[htbp]\centering
\footnotesize
\begin{threeparttable}
\def\sym#1{\ifmmode^{#1}\else\(^{#1}\)\fi}
\caption{Pre-Market Emotions, Investor Types and Price Movements\label{tab:pre_market_types}}
\begin{tabular}{l*{6}{c}}
\hline\hline
                    &\multicolumn{1}{c}{(1)}&\multicolumn{1}{c}{(2)}&\multicolumn{1}{c}{(3)}&\multicolumn{1}{c}{(4)}&\multicolumn{1}{c}{(5)}&\multicolumn{1}{c}{(6)}\\
                  & \multicolumn{2}{c}{Trading Experience} & \multicolumn{2}{c}{Trading Approach} & \multicolumn{2}{c}{Investment Horizon}  \\
                 &  Non-Professional & Professional & Fundamental & Technical & Short-Term & Long-Term \\ \hline
   &        &        &        &        &        &           \\[\dimexpr-\normalbaselineskip+2pt]
StockTwits Sentiment           &      0.0023\sym{***}&      0.0023\sym{***}&      0.0020\sym{***}&      0.0019\sym{***}&      0.0023\sym{***}&      0.0022\sym{***}\\
                    &    (0.0002)         &    (0.0002)         &    (0.0002)         &    (0.0002)         &    (0.0002)         &    (0.0002)         \\
EmTract Valence             &      0.0017\sym{***}&      0.0009\sym{***}&      0.0015\sym{***}&      0.0017\sym{***}&      0.0017\sym{***}&      0.0016\sym{***}\\
                    &    (0.0003)         &    (0.0003)         &    (0.0003)         &    (0.0003)         &    (0.0003)         &    (0.0003)         \\
   &        &        &        &        &        &           \\[\dimexpr-\normalbaselineskip+2pt]

\hline
   &        &        &        &        &        &           \\[\dimexpr-\normalbaselineskip+2pt]
Observations        & 369159       & 384115       & 377388       & 384337       & 395470       & 410174       \\
$R^2$               &      0.1128         &      0.1054         &      0.1066         &      0.1061         &      0.1051         &      0.1074         \\
\hline\hline
\end{tabular}
\begin{tablenotes}
\footnotesize 
\item Notes: This table considers the relationship between pre-market emotions, StockTwits sentiment, user characteristics and daily price movements. The dependent variable is daily returns, computed as the difference between the closing and the opening price, divided by the opening price. All specifications include firm and date fixed effects, close-open returns, lag open-close returns, past 20 trading days return and volatility (as defined in Variable Definitions). Robust standard errors clustered at the industry and date levels are in parentheses. We use the Fama-French 12-industry classification. \sym{*} \(p<0.10\), \sym{**} \(p<0.05\), \sym{***} \(p<0.01\). Continuous variables winsorized at the 0.1\% and 99.9\% level to mitigate the impact of outliers. We report the within-firm, detrended (demeaned) standard deviation of the dependent variable. 
\end{tablenotes}
\end{threeparttable}
\end{sidewaystable}